\documentclass[twocolumn,trackchanges]{aastex701}

\usepackage{subcaption}
\usepackage{amsmath}

\DeclareMathOperator\erf{erf}
\DeclareMathOperator*\argmin{argmin}

\begin{document}

\title{Polarization Angle Swings in Blazars Detected in the Millimeter-wave with the South Pole Telescope}

\author[0000-0001-5396-2850]{A.~Simpson}
\affiliation{Department of Astronomy and Astrophysics, University of Chicago, 5640 South Ellis Avenue, Chicago, IL, 60637, USA}
\affiliation{Kavli Institute for Cosmological Physics, University of Chicago, 5640 South Ellis Avenue, Chicago, IL, 60637, USA}
\email{simpsa@uchicago.edu}

\author[0009-0003-3245-3979]{E.~Anderes}
\affiliation{Department of Statistics, University of California, One Shields Avenue, Davis, CA 95616, USA}
\email{ebanderes@ucdavis.edu}

\author[0000-0002-4435-4623]{A.~J.~Anderson}
\affiliation{Fermi National Accelerator Laboratory, MS209, P.O. Box 500, Batavia, IL, 60510, USA}
\affiliation{Kavli Institute for Cosmological Physics, University of Chicago, 5640 South Ellis Avenue, Chicago, IL, 60637, USA}
\affiliation{Department of Astronomy and Astrophysics, University of Chicago, 5640 South Ellis Avenue, Chicago, IL, 60637, USA}
\email{adama@fnal.gov}

\author{B.~Ansarinejad}
\affiliation{School of Physics, University of Melbourne, Parkville, VIC 3010, Australia}
\email{behzad.ansarinejad@gmail.com}

\author[0000-0002-0517-9842]{M.~Archipley}
\affiliation{Department of Astronomy and Astrophysics, University of Chicago, 5640 South Ellis Avenue, Chicago, IL, 60637, USA}
\affiliation{Kavli Institute for Cosmological Physics, University of Chicago, 5640 South Ellis Avenue, Chicago, IL, 60637, USA}
\email{archipleym@uchicago.edu}

\author[0000-0001-6899-1873]{L.~Balkenhol}
\affiliation{Sorbonne Universit\'e, CNRS, UMR 7095, Institut d'Astrophysique de Paris, 98 bis bd Arago, 75014 Paris, France}
\email{lennart.balkenhol@iap.fr}

\author[0000-0002-1623-5651]{D.~R.~Barron}
\affiliation{Department of Physics and Astronomy, University of New Mexico, Albuquerque, NM, 87131, USA}
\email{barrondarcy@gmail.com}

\author[0000-0001-9103-9354]{P.~S.~Barry}
\affiliation{School of Physics and Astronomy, Cardiff University, Cardiff CF24 3YB, United Kingdom}
\email{BarryP2@cardiff.ac.uk}

\author{K.~Benabed}
\affiliation{Sorbonne Universit\'e, CNRS, UMR 7095, Institut d'Astrophysique de Paris, 98 bis bd Arago, 75014 Paris, France}
\email{benabed@iap.fr}

\author[0000-0001-5868-0748]{A.~N.~Bender}
\affiliation{High-Energy Physics Division, Argonne National Laboratory, 9700 South Cass Avenue, Lemont, IL, 60439, USA}
\affiliation{Kavli Institute for Cosmological Physics, University of Chicago, 5640 South Ellis Avenue, Chicago, IL, 60637, USA}
\affiliation{Department of Astronomy and Astrophysics, University of Chicago, 5640 South Ellis Avenue, Chicago, IL, 60637, USA}
\email{abender@anl.gov}

\author[0000-0002-5108-6823]{B.~A.~Benson}
\affiliation{Fermi National Accelerator Laboratory, MS209, P.O. Box 500, Batavia, IL, 60510, USA}
\affiliation{Kavli Institute for Cosmological Physics, University of Chicago, 5640 South Ellis Avenue, Chicago, IL, 60637, USA}
\affiliation{Department of Astronomy and Astrophysics, University of Chicago, 5640 South Ellis Avenue, Chicago, IL, 60637, USA}
\email{bbenson@astro.uchicago.edu}

\author[0000-0003-4847-3483]{F.~Bianchini}
\affiliation{Kavli Institute for Particle Astrophysics and Cosmology, Stanford University, 452 Lomita Mall, Stanford, CA, 94305, USA}
\affiliation{Department of Physics, Stanford University, 382 Via Pueblo Mall, Stanford, CA, 94305, USA}
\affiliation{SLAC National Accelerator Laboratory, 2575 Sand Hill Road, Menlo Park, CA, 94025, USA}
\email{fbianc@stanford.edu}

\author[0000-0001-7665-5079]{L.~E.~Bleem}
\affiliation{High-Energy Physics Division, Argonne National Laboratory, 9700 South Cass Avenue, Lemont, IL, 60439, USA}
\affiliation{Kavli Institute for Cosmological Physics, University of Chicago, 5640 South Ellis Avenue, Chicago, IL, 60637, USA}
\affiliation{Department of Astronomy and Astrophysics, University of Chicago, 5640 South Ellis Avenue, Chicago, IL, 60637, USA}
\email{lbleem@anl.gov}

\author[0000-0002-4900-805X]{S.~Bocquet}
\affiliation{University Observatory, Faculty of Physics, LMU Munich, Scheinerstr.~1, 81679 Munich, Germany}
\email{sebastian.bocquet@physik.lmu.de}

\author[0000-0002-8051-2924]{F.~R.~Bouchet}
\affiliation{Sorbonne Universit\'e, CNRS, UMR 7095, Institut d'Astrophysique de Paris, 98 bis bd Arago, 75014 Paris, France}
\email{bouchet@iap.fr}

\author[0000-0003-3483-8461]{E.~Camphuis}
\affiliation{Sorbonne Universit\'e, CNRS, UMR 7095, Institut d'Astrophysique de Paris, 98 bis bd Arago, 75014 Paris, France}
\email{etienne.camphuis@iap.fr}

\author{M.~G.~Campitiello}
\affiliation{High-Energy Physics Division, Argonne National Laboratory, 9700 South Cass Avenue, Lemont, IL, 60439, USA}
\email{mcampitiello@anl.gov}

\author[0000-0002-2044-7665]{J.~E.~Carlstrom}
\affiliation{Kavli Institute for Cosmological Physics, University of Chicago, 5640 South Ellis Avenue, Chicago, IL, 60637, USA}
\affiliation{Enrico Fermi Institute, University of Chicago, 5640 South Ellis Avenue, Chicago, IL, 60637, USA}
\affiliation{Department of Physics, University of Chicago, 5640 South Ellis Avenue, Chicago, IL, 60637, USA}
\affiliation{High-Energy Physics Division, Argonne National Laboratory, 9700 South Cass Avenue, Lemont, IL, 60439, USA}
\affiliation{Department of Astronomy and Astrophysics, University of Chicago, 5640 South Ellis Avenue, Chicago, IL, 60637, USA}
\email{jc@astro.uchicago.edu}

\author[0000-0002-5751-1392]{J.~Carron}
\affiliation{Istituto ricerche solari Aldo e Cele Dacc\`o (IRSOL), Faculty of Informatics, Universit\`a della Svizzera italiana, 6605 Locarno, Switzerland}
\affiliation{Universit\'e de Gen\`eve, D\'epartement de Physique Th\'eorique, 24 Quai Ansermet, CH-1211 Gen\`eve 4, Switzerland}
\email{to.jcarron@gmail.com}

\author{C.~L.~Chang}
\affiliation{High-Energy Physics Division, Argonne National Laboratory, 9700 South Cass Avenue, Lemont, IL, 60439, USA}
\affiliation{Kavli Institute for Cosmological Physics, University of Chicago, 5640 South Ellis Avenue, Chicago, IL, 60637, USA}
\affiliation{Department of Astronomy and Astrophysics, University of Chicago, 5640 South Ellis Avenue, Chicago, IL, 60637, USA}
\email{clchang@uchicago.edu}

\author[0000-0002-5397-9035]{P.~M.~Chichura}
\affiliation{Department of Physics, University of Chicago, 5640 South Ellis Avenue, Chicago, IL, 60637, USA}
\affiliation{Kavli Institute for Cosmological Physics, University of Chicago, 5640 South Ellis Avenue, Chicago, IL, 60637, USA}
\email{pchichura@uchicago.edu}

\author{A.~Chokshi}
\affiliation{Department of Astronomy and Astrophysics, University of Chicago, 5640 South Ellis Avenue, Chicago, IL, 60637, USA}
\email{aman.chokshi@mcgill.ca}

\author[0000-0002-3091-8790]{T.-L.~Chou}
\affiliation{Department of Astronomy and Astrophysics, University of Chicago, 5640 South Ellis Avenue, Chicago, IL, 60637, USA}
\affiliation{Kavli Institute for Cosmological Physics, University of Chicago, 5640 South Ellis Avenue, Chicago, IL, 60637, USA}
\affiliation{National Taiwan University, No. 1, Sec. 4, Roosevelt Road, Taipei 106319, Taiwan}
\email{tlchou@uchicago.edu}

\author[0000-0002-2707-1672]{A.~Coerver}
\affiliation{Department of Physics, University of California, Berkeley, CA, 94720, USA}
\email{acoerver@berkeley.edu}

\author[0000-0001-9000-5013]{T.~M.~Crawford}
\affiliation{Department of Astronomy and Astrophysics, University of Chicago, 5640 South Ellis Avenue, Chicago, IL, 60637, USA}
\affiliation{Kavli Institute for Cosmological Physics, University of Chicago, 5640 South Ellis Avenue, Chicago, IL, 60637, USA}
\email{tmcrawfo@uchicago.edu}

\author[0000-0002-3760-2086]{C.~Daley}
\affiliation{Universit\'e Paris-Saclay, Universit\'e Paris Cit\'e, CEA, CNRS, AIM, 91191, Gif-sur-Yvette, France}
\affiliation{Department of Astronomy, University of Illinois Urbana-Champaign, 1002 West Green Street, Urbana, IL, 61801, USA}
\email{cailmd2@illinois.edu}

\author[0000-0001-5105-9473]{T.~de~Haan}
\affiliation{High Energy Accelerator Research Organization (KEK), Tsukuba, Ibaraki 305-0801, Japan}
\email{tijmen.dehaan@gmail.com}

\author{K.~R.~Dibert}
\affiliation{Department of Astronomy and Astrophysics, University of Chicago, 5640 South Ellis Avenue, Chicago, IL, 60637, USA}
\affiliation{Kavli Institute for Cosmological Physics, University of Chicago, 5640 South Ellis Avenue, Chicago, IL, 60637, USA}
\email{krdibert@uchicago.edu}

\author{M.~A.~Dobbs}
\affiliation{Department of Physics and McGill Space Institute, McGill University, 3600 Rue University, Montreal, Quebec H3A 2T8, Canada}
\affiliation{Canadian Institute for Advanced Research, CIFAR Program in Gravity and the Extreme Universe, Toronto, ON, M5G 1Z8, Canada}
\email{matt.dobbs@mcgill.ca}

\author{M.~Doohan}
\affiliation{School of Physics, University of Melbourne, Parkville, VIC 3010, Australia}
\email{mdoohan@student.unimelb.edu.au}

\author[0000-0002-9962-2058]{D.~Dutcher}
\affiliation{Joseph Henry Laboratories of Physics, Jadwin Hall, Princeton University, Princeton, NJ 08544, USA}
\email{ddutcher@uchicago.edu}

\author{C.~Feng}
\affiliation{Department of Astronomy, University of Science and Technology of China, Hefei 230026, China}
\affiliation{School of Astronomy and Space Science, University of Science and Technology of China, Hefei 230026}
\affiliation{Department of Physics, University of Illinois Urbana-Champaign, 1110 West Green Street, Urbana, IL, 61801, USA}
\email{changfeng@ustc.edu.cn}

\author[0000-0002-4928-8813]{K.~R.~Ferguson}
\affiliation{Department of Physics and Astronomy, University of California, Los Angeles, CA, 90095, USA}
\affiliation{Department of Physics and Astronomy, Michigan State University, East Lansing, MI 48824, USA}
\email{kferguson@physics.ucla.edu}

\author[0000-0002-7130-7099]{N.~C.~Ferree}
\affiliation{California Institute of Technology, 1200 East California Boulevard., Pasadena, CA, 91125, USA}
\affiliation{Kavli Institute for Particle Astrophysics and Cosmology, Stanford University, 452 Lomita Mall, Stanford, CA, 94305, USA}
\affiliation{Department of Physics, Stanford University, 382 Via Pueblo Mall, Stanford, CA, 94305, USA}
\email{nferree@stanford.edu}

\author{K.~Fichman}
\affiliation{Department of Physics, University of Chicago, 5640 South Ellis Avenue, Chicago, IL, 60637, USA}
\affiliation{Kavli Institute for Cosmological Physics, University of Chicago, 5640 South Ellis Avenue, Chicago, IL, 60637, USA}
\email{kfichman@uchicago.edu}

\author[0000-0002-7145-1824]{A.~Foster}
\affiliation{Joseph Henry Laboratories of Physics, Jadwin Hall, Princeton University, Princeton, NJ 08544, USA}
\email{axf295@case.edu}

\author{S.~Galli}
\affiliation{Sorbonne Universit\'e, CNRS, UMR 7095, Institut d'Astrophysique de Paris, 98 bis bd Arago, 75014 Paris, France}
\email{gallis@iap.fr}

\author{A.~E.~Gambrel}
\affiliation{Kavli Institute for Cosmological Physics, University of Chicago, 5640 South Ellis Avenue, Chicago, IL, 60637, USA}
\email{anne.gambrel@gmail.com}

\author{A.~K.~Gao}
\affiliation{Department of Physics, University of Illinois Urbana-Champaign, 1110 West Green Street, Urbana, IL, 61801, USA}
\email{akgao2@illinois.edu}

\author{F.~Ge}
\affiliation{California Institute of Technology, 1200 East California Boulevard., Pasadena, CA, 91125, USA}
\affiliation{Kavli Institute for Particle Astrophysics and Cosmology, Stanford University, 452 Lomita Mall, Stanford, CA, 94305, USA}
\affiliation{Department of Physics, Stanford University, 382 Via Pueblo Mall, Stanford, CA, 94305, USA}
\affiliation{Department of Physics \& Astronomy, University of California, One Shields Avenue, Davis, CA 95616, USA}
\email{fge@ucdavis.edu}

\author[0000-0001-7593-3962]{F.~Guidi}
\affiliation{Department of Physics \& Astronomy, University of California, One Shields Avenue, Davis, CA 95616, USA}
\affiliation{Sorbonne Universit\'e, CNRS, UMR 7095, Institut d'Astrophysique de Paris, 98 bis bd Arago, 75014 Paris, France}
\email{federica.guidi@iap.fr}

\author{S.~Guns}
\affiliation{Department of Physics, University of California, Berkeley, CA, 94720, USA}
\email{sguns@berkeley.edu}

\author{N.~W.~Halverson}
\affiliation{CASA, Department of Astrophysical and Planetary Sciences, University of Colorado, Boulder, CO, 80309, USA }
\affiliation{Department of Physics, University of Colorado, Boulder, CO, 80309, USA}
\email{nils.halverson@colorado.edu}

\author[0000-0003-1690-6678]{A.~D.~Hincks}
\affiliation{Dunlap Institute for Astronomy \& Astrophysics, University of Toronto, 50 St. George Street, Toronto, ON, M5S 3H4, Canada}
\affiliation{Specola Vaticana (Vatican Observatory), V-00120 Vatican City, Vatican City State}
\email{adam.hincks@utoronto.ca}

\author[0000-0003-1880-2733]{E.~Hivon}
\affiliation{Sorbonne Universit\'e, CNRS, UMR 7095, Institut d'Astrophysique de Paris, 98 bis bd Arago, 75014 Paris, France}
\email{hivon@iap.fr}

\author[0000-0002-0463-6394]{G.~P.~Holder}
\affiliation{Department of Physics, University of Illinois Urbana-Champaign, 1110 West Green Street, Urbana, IL, 61801, USA}
\email{gholder@illinois.edu}

\author{W.~L.~Holzapfel}
\affiliation{Department of Physics, University of California, Berkeley, CA, 94720, USA}
\email{swlh@cosmology.berkeley.edu}

\author{J.~C.~Hood}
\affiliation{Kavli Institute for Cosmological Physics, University of Chicago, 5640 South Ellis Avenue, Chicago, IL, 60637, USA}
\email{hoodjc@uchicago.edu}

\author{A.~Hryciuk}
\affiliation{Department of Physics, University of Chicago, 5640 South Ellis Avenue, Chicago, IL, 60637, USA}
\affiliation{Kavli Institute for Cosmological Physics, University of Chicago, 5640 South Ellis Avenue, Chicago, IL, 60637, USA}
\email{hryciuk@uchicago.edu}

\author[0000-0003-3595-0359]{N.~Huang}
\affiliation{Department of Physics, University of California, Berkeley, CA, 94720, USA}
\email{nhuang@alumni.princeton.edu}

\author{E.~J\"arvel\"a}
\affiliation{Department of Physics \& Astronomy, Box 41051, Texas Tech University, Lubbock TX 79409-1051, USA}
\email{EMAIL}

\author{T.~Jhaveri}
\affiliation{Department of Astronomy and Astrophysics, University of Chicago, 5640 South Ellis Avenue, Chicago, IL, 60637, USA}
\affiliation{Kavli Institute for Cosmological Physics, University of Chicago, 5640 South Ellis Avenue, Chicago, IL, 60637, USA}
\email{tanishaj@uchicago.edu}

\author{F.~K\'eruzor\'e}
\affiliation{High-Energy Physics Division, Argonne National Laboratory, 9700 South Cass Avenue, Lemont, IL, 60439, USA}
\email{fkeruzore@anl.gov}

\author[0000-0002-8388-4950]{A.~R.~Khalife}
\affiliation{Sorbonne Universit\'e, CNRS, UMR 7095, Institut d'Astrophysique de Paris, 98 bis bd Arago, 75014 Paris, France}
\email{ridakhal@iap.fr}

\author{L.~Knox}
\affiliation{Department of Physics \& Astronomy, University of California, One Shields Avenue, Davis, CA 95616, USA}
\email{lknox@ucdavis.edu}

\author{K.~Kornoelje}
\affiliation{Department of Astronomy and Astrophysics, University of Chicago, 5640 South Ellis Avenue, Chicago, IL, 60637, USA}
\affiliation{Kavli Institute for Cosmological Physics, University of Chicago, 5640 South Ellis Avenue, Chicago, IL, 60637, USA}
\affiliation{High-Energy Physics Division, Argonne National Laboratory, 9700 South Cass Avenue, Lemont, IL, 60439, USA}
\email{knk@uchicago.edu}

\author{C.-L.~Kuo}
\affiliation{Kavli Institute for Particle Astrophysics and Cosmology, Stanford University, 452 Lomita Mall, Stanford, CA, 94305, USA}
\affiliation{Department of Physics, Stanford University, 382 Via Pueblo Mall, Stanford, CA, 94305, USA}
\affiliation{SLAC National Accelerator Laboratory, 2575 Sand Hill Road, Menlo Park, CA, 94025, USA}
\email{clkuo@stanford.edu}

\author{K.~Levy}
\affiliation{School of Physics, University of Melbourne, Parkville, VIC 3010, Australia}
\email{kevin.levy@student.unimelb.edu.au}

\author[0000-0002-4820-1122]{Y.~Li}
\affiliation{Kavli Institute for Cosmological Physics, University of Chicago, 5640 South Ellis Avenue, Chicago, IL, 60637, USA}
\email{yunyangl@uchicago.edu}

\author[0000-0002-4747-4276]{A.~E.~Lowitz}
\affiliation{Kavli Institute for Cosmological Physics, University of Chicago, 5640 South Ellis Avenue, Chicago, IL, 60637, USA}
\email{lowitz@arizona.edu}

\author{C.~Lu}
\affiliation{Department of Physics, University of Illinois Urbana-Champaign, 1110 West Green Street, Urbana, IL, 61801, USA}
\email{chunyul3@illinois.edu}

\author[0009-0004-3143-1708]{G.~P.~Lynch}
\affiliation{Department of Physics \& Astronomy, University of California, One Shields Avenue, Davis, CA 95616, USA}
\email{gplynch@ucdavis.edu}

\author[0000-0003-2622-6895]{X.~Ma}
\affiliation{Kavli Institute for Astronomy and Astrophysics, Peking University, Beijing 100871, China}
\affiliation{Department of Astronomy, School of Physics, Peking University, Beijing 100871, China}
\email{aonexy@gmail.com}

\author[0000-0003-0976-4755]{T.~J.~Maccarone}
\affiliation{Department of Physics \& Astronomy, Box 41051, Texas Tech University, Lubbock TX 79409-1051, USA}
\email{thomas.maccarone@ttu.edu}

\author[0009-0003-8076-283X]{G.~Madejski}
\affiliation{Kavli Institute for Particle Astrophysics and Cosmology and SLAC National Accelerator Laboratory, Stanford University, Menlo Park, California 94025, USA}
\email{madejski@slac.stanford.edu}

\author[0000-0002-4617-9320]{A.~S.~Maniyar}
\affiliation{Kavli Institute for Particle Astrophysics and Cosmology, Stanford University, 452 Lomita Mall, Stanford, CA, 94305, USA}
\affiliation{Department of Physics, Stanford University, 382 Via Pueblo Mall, Stanford, CA, 94305, USA}
\affiliation{SLAC National Accelerator Laboratory, 2575 Sand Hill Road, Menlo Park, CA, 94025, USA}
\email{abhishek.maniyar@lapth.cnrs.fr}

\author{E.~S.~Martsen}
\affiliation{Department of Astronomy and Astrophysics, University of Chicago, 5640 South Ellis Avenue, Chicago, IL, 60637, USA}
\affiliation{Kavli Institute for Cosmological Physics, University of Chicago, 5640 South Ellis Avenue, Chicago, IL, 60637, USA}
\email{emartsen@uchicago.edu}

\author{F.~Menanteau}
\affiliation{Department of Astronomy, University of Illinois Urbana-Champaign, 1002 West Green Street, Urbana, IL, 61801, USA}
\affiliation{Center for AstroPhysical Surveys, National Center for Supercomputing Applications, Urbana, IL, 61801, USA}
\email{felipe@illinois.edu}

\author[0000-0001-7317-0551]{M.~Millea}
\affiliation{Department of Physics, University of California, Berkeley, CA, 94720, USA}
\email{mariusmillea@gmail.com}

\author{J.~Montgomery}
\affiliation{Department of Physics and McGill Space Institute, McGill University, 3600 Rue University, Montreal, Quebec H3A 2T8, Canada}
\email{joshua.j.montgomery@gmail.com}

\author{Y.~Nakato}
\affiliation{Department of Physics, Stanford University, 382 Via Pueblo Mall, Stanford, CA, 94305, USA}
\email{yukanaka@stanford.edu}

\author{T.~Natoli}
\affiliation{Kavli Institute for Cosmological Physics, University of Chicago, 5640 South Ellis Avenue, Chicago, IL, 60637, USA}
\email{tnatoli2@gmail.com}

\author[0000-0003-0170-5638]{A.~Ouellette}
\affiliation{Department of Physics, University of Illinois Urbana-Champaign, 1110 West Green Street, Urbana, IL, 61801, USA}
\email{aaronjo2@illinois.edu}

\author[0000-0002-6164-9861]{Z.~Pan}
\affiliation{High-Energy Physics Division, Argonne National Laboratory, 9700 South Cass Avenue, Lemont, IL, 60439, USA}
\affiliation{Kavli Institute for Cosmological Physics, University of Chicago, 5640 South Ellis Avenue, Chicago, IL, 60637, USA}
\affiliation{Department of Physics, University of Chicago, 5640 South Ellis Avenue, Chicago, IL, 60637, USA}
\email{panz@uchicago.edu}

\author[0000-0001-7946-557X]{K.~A.~Phadke}
\affiliation{Department of Astronomy, University of Illinois Urbana-Champaign, 1002 West Green Street, Urbana, IL, 61801, USA}
\affiliation{Center for AstroPhysical Surveys, National Center for Supercomputing Applications, Urbana, IL, 61801, USA}
\affiliation{NSF-Simons AI Institute for the Sky (SkAI), 172 E. Chestnut St., Chicago, IL 60611, USA}
\email{kphadke2@illinois.edu}

\author{A.~W.~Pollak}
\affiliation{Department of Astronomy and Astrophysics, University of Chicago, 5640 South Ellis Avenue, Chicago, IL, 60637, USA}
\email{alexander.pollak.87@gmail.com}

\author{K.~Prabhu}
\affiliation{Department of Physics \& Astronomy, University of California, One Shields Avenue, Davis, CA 95616, USA}
\email{kprabhu@ucdavis.edu}

\author[0009-0002-2589-5501]{W.~Quan}
\affiliation{High-Energy Physics Division, Argonne National Laboratory, 9700 South Cass Avenue, Lemont, IL, 60439, USA}
\affiliation{Department of Physics, University of Chicago, 5640 South Ellis Avenue, Chicago, IL, 60637, USA}
\affiliation{Kavli Institute for Cosmological Physics, University of Chicago, 5640 South Ellis Avenue, Chicago, IL, 60637, USA}
\email{weiquan@uchicago.edu}

\author{M.~Rahimi}
\affiliation{School of Physics, University of Melbourne, Parkville, VIC 3010, Australia}
\email{mahsa.rahimi@unimelb.edu.au}

\author[0000-0003-3953-1776]{A.~Rahlin}
\affiliation{Department of Astronomy and Astrophysics, University of Chicago, 5640 South Ellis Avenue, Chicago, IL, 60637, USA}
\affiliation{Kavli Institute for Cosmological Physics, University of Chicago, 5640 South Ellis Avenue, Chicago, IL, 60637, USA}
\email{arahlin@uchicago.edu}

\author[0000-0003-2226-9169]{C.~L.~Reichardt}
\affiliation{School of Physics, University of Melbourne, Parkville, VIC 3010, Australia}
\email{clreichardt@gmail.com}

\author{M.~Rouble}
\affiliation{Department of Physics and McGill Space Institute, McGill University, 3600 Rue University, Montreal, Quebec H3A 2T8, Canada}
\email{maclean.rouble@mail.mcgill.ca}

\author{J.~E.~Ruhl}
\affiliation{Department of Physics, Case Western Reserve University, Cleveland, OH, 44106, USA}
\email{ruhl@case.edu}

\author[0000-0001-5755-5865]{A.~C.~Silva~Oliveira}
\affiliation{California Institute of Technology, 1200 East California Boulevard., Pasadena, CA, 91125, USA}
\affiliation{Kavli Institute for Particle Astrophysics and Cosmology, Stanford University, 452 Lomita Mall, Stanford, CA, 94305, USA}
\affiliation{Department of Physics, Stanford University, 382 Via Pueblo Mall, Stanford, CA, 94305, USA}
\email{anaoliv@stanford.edu}

\author[0000-0001-6155-5315]{J.~A.~Sobrin}
\affiliation{Fermi National Accelerator Laboratory, MS209, P.O. Box 500, Batavia, IL, 60510, USA}
\affiliation{Kavli Institute for Cosmological Physics, University of Chicago, 5640 South Ellis Avenue, Chicago, IL, 60637, USA}
\email{jsobrin@fnal.gov}

\author{A.~A.~Stark}
\affiliation{Center for Astrophysics \textbar{} Harvard \& Smithsonian, 60 Garden Street, Cambridge, MA, 02138, USA}
\email{astark@cfa.harvard.edu}

\author[0000-0002-2077-6004]{C.~Tandoi}
\affiliation{Department of Astronomy, University of Illinois Urbana-Champaign, 1002 West Green Street, Urbana, IL, 61801, USA}
\email{ctandoi2@illinois.edu}

\author{C.~Trendafilova}
\affiliation{Center for AstroPhysical Surveys, National Center for Supercomputing Applications, Urbana, IL, 61801, USA}
\email{ctrendaf@illinois.edu}

\author[0000-0001-7192-3871]{J.~D.~Vieira}
\affiliation{Department of Astronomy, University of Illinois Urbana-Champaign, 1002 West Green Street, Urbana, IL, 61801, USA}
\affiliation{Department of Physics, University of Illinois Urbana-Champaign, 1110 West Green Street, Urbana, IL, 61801, USA}
\affiliation{Center for AstroPhysical Surveys, National Center for Supercomputing Applications, Urbana, IL, 61801, USA}
\email{jvieira@illinois.edu}

\author[0000-0002-4528-9886]{A.~G.~Vieregg}
\affiliation{Kavli Institute for Cosmological Physics, University of Chicago, 5640 South Ellis Avenue, Chicago, IL, 60637, USA}
\affiliation{Department of Astronomy and Astrophysics, University of Chicago, 5640 South Ellis Avenue, Chicago, IL, 60637, USA}
\affiliation{Enrico Fermi Institute, University of Chicago, 5640 South Ellis Avenue, Chicago, IL, 60637, USA}
\affiliation{Department of Physics, University of Chicago, 5640 South Ellis Avenue, Chicago, IL, 60637, USA}
\email{avieregg@kicp.uchicago.edu}

\author[0009-0009-3168-092X]{A.~Vitrier}
\affiliation{Sorbonne Universit\'e, CNRS, UMR 7095, Institut d'Astrophysique de Paris, 98 bis bd Arago, 75014 Paris, France}
\email{aline.vitrier@iap.fr}

\author{Y.~Wan}
\affiliation{Department of Astronomy, University of Illinois Urbana-Champaign, 1002 West Green Street, Urbana, IL, 61801, USA}
\affiliation{Center for AstroPhysical Surveys, National Center for Supercomputing Applications, Urbana, IL, 61801, USA}
\email{yujiew2@illinois.edu}

\author[0000-0002-3157-0407]{N.~Whitehorn}
\affiliation{Department of Physics and Astronomy, Michigan State University, East Lansing, MI 48824, USA}
\email{nathanw@msu.edu}

\author[0000-0001-5411-6920]{W.~L.~K.~Wu}
\affiliation{California Institute of Technology, 1200 East California Boulevard., Pasadena, CA, 91125, USA}
\affiliation{Kavli Institute for Particle Astrophysics and Cosmology, Stanford University, 452 Lomita Mall, Stanford, CA, 94305, USA}
\affiliation{SLAC National Accelerator Laboratory, 2575 Sand Hill Road, Menlo Park, CA, 94025, USA}
\email{kimwuu@gmail.com}

\author{M.~R.~Young}
\affiliation{Fermi National Accelerator Laboratory, MS209, P.O. Box 500, Batavia, IL, 60510, USA}
\affiliation{Kavli Institute for Cosmological Physics, University of Chicago, 5640 South Ellis Avenue, Chicago, IL, 60637, USA}
\email{mattyoungofficial@gmail.com}

\author{J.~A.~Zebrowski}
\affiliation{Kavli Institute for Cosmological Physics, University of Chicago, 5640 South Ellis Avenue, Chicago, IL, 60637, USA}
\affiliation{Department of Astronomy and Astrophysics, University of Chicago, 5640 South Ellis Avenue, Chicago, IL, 60637, USA}
\affiliation{Fermi National Accelerator Laboratory, MS209, P.O. Box 500, Batavia, IL, 60510, USA}
\email{j.z@uchicago.edu}

\collaboration{all}{The South Pole Telescope Collaboration}

\begin{abstract}

    We present the first systematic search for electric vector position angle (EVPA) swings in the millimeter-wave (mm-wave) emission of blazars, using five years of observations from the South Pole Telescope SPT-3G camera at 95 and 150~GHz, and investigate their connection to gamma-ray flares.
    Of the 168 bright sources in the $\sim$1500~deg$^2$ SPT-3G Main Field, eight have sufficient polarization signal-to-noise for reliable EVPA measurement, four of which have continuous \textit{Fermi} LAT gamma-ray detections.
    We detect EVPA swings in all four gamma-ray-active blazars and in none of the remaining four, consistent with the established connection between EVPA swings and high-energy emission seen at optical wavelengths.
    The observed swing amplitudes and rotation rates are smaller than those found in optical studies, consistent with mm-wave EVPA variability being slower than at shorter wavelengths.
    Random walk simulations of the polarization angle using a multi-cell model fail to reproduce both the number, amplitude, and duration of the observed swings, suggesting that this mechanism is insufficient to explain the observed swing population.
    Analysis of EVPA variability on one-week timescales is consistent with the anti-correlation between polarization degree and EVPA rotation rate previously observed at optical wavelengths.
    Of the 14 detected swings, eight are within 30 days of a gamma-ray flare. While many individual swing-flare associations are found to have a very low probability of happening by chance, the full ensembles of 95 and 150~GHz swing time lags with respect to gamma-ray flares are found to be consistent with random coincidence.

\end{abstract}


\section{Introduction}
\label{Sec:Introduction}

Blazars, a class of active galactic nuclei (AGN), are characterized by relativistic jets aligned at a narrow angle relative to the observer's line of sight \citep{urry95}.
The relativistic velocity of the jet enhances emission through Doppler boosting when directed along the observed line of sight.
Blazars radiate across the entire observed electromagnetic spectrum, ranging from radio waves to GeV and even TeV gamma-rays, characterized by a double-peaked spectral energy distribution (SED, \citealt{fossati98}).
The lower-frequency SED peak is dominated by synchrotron radiation from the jet, produced by relativistic electrons, which is expected to be highly polarized in the optically thin regime \citep[e.g.,][]{longair11} if the magnetic field of the jet is well-ordered.
The emission from blazars exhibits significant variability across all observed timescales and frequencies \citep[e.g.,][]{blandford19}.
This variability extends beyond the brightness of the emission, with significant fluctuations observed in two polarization characteristics: polarization degree (PD, $\Pi$) and electric vector polarization angle (EVPA, $\chi$).
The polarization characteristics are dependent on the properties of the emission region and the configuration of the magnetic field within the jet.

Variable polarized emission of blazars has been detected at radio \citep[e.g.,][]{darcangelo09}, millimeter \citep[mm, e.g.,][]{agudo22}, optical \citep[e.g.,][]{dominici08}, and more recently X-ray \citep[e.g.,][]{liodakis22} wavelengths.
Prior observations of the variability of polarized emission of blazars show that it can vary strongly on single-day, or even hour, time-scales \citep[e.g.,][]{wagner95}.
While the POLAMI program \citep{agudo22} has monitored the mm-wave polarization of a sample of blazars over an extended period, few blazars have been monitored at the cadence required to characterize their polarization variability on day time-scales.

Recent observations at optical wavelengths of large EVPA changes in blazars \citep[e.g.,][]{blinov15}, which we refer to in this work as ``EVPA swings,'' have sparked interest in studying this particular blazar phenomenon.
In 2009, a large change in EVPA and PD at optical wavelengths was observed to be coincident with a gamma-ray flare in the blazar 3C 279 \citep{abdo10a}.
If these events are physically linked, they may provide crucial insights into the physical mechanisms underlying gamma-ray activity and the spatial origin of gamma-ray emission in the jet.
Subsequent investigations by the RoboPol collaboration \citep{blinov15}, in which a large unbiased population of blazars was monitored, concluded that some optical EVPA swings are indeed physically linked with gamma-ray flares.

Models proposed to explain the EVPA variability and its potential connection with gamma-ray flares can be categorized into two classes: stochastic and deterministic.
Stochastic models \citep[e.g.,][]{jones88, marscher14} characterize the jet as comprising numerous cells, whose random fluctuations induce significant EVPA changes.
Previous studies have shown that such models are capable of producing large EVPA variations, including apparent EVPA swings.
In contrast, deterministic models link the observed variability to specific events or the underlying jet structure.
These models include, but are not limited to, relativistic aberration \citep[e.g.,][]{abdo10a}, alterations to the structure of the magnetic field \citep[e.g.,][]{zhang15, zhang16}, the superposition of several emission components with different polarization characteristics \citep{holmes84}, and non-axisymmetric jet structure \citep[e.g.,][]{marscher08}.

In recent years, new cosmic microwave background (CMB) experiments with improved angular resolution have been recognized for their ability to serve as mm-wave transient monitors, as a byproduct of their observation strategy.
Two CMB experiments that have recently demonstrated major promise in the area of transient and variable sources are the South Pole Telescope (SPT, \citealt{carlstrom11}), and the Atacama Cosmology Telescope (ACT, \citealt{fowler07,swetz11, thornton16}).
Recent examples of SPT and ACT transient and variable source science include observations of stellar flares \citep{naess21, tandoi24}, detection of galactic and extragalactic transients \citep{guns21, Hervias_Caimapo24, biermann25}, detection of flares from accreting white dwarf systems \citep{wan26}, supermassive black hole binary candidate observations \citep{hincks26}, and a catalog of AGN light curves \citep{hood26}.
In this work we primarily focus on data from the SPT, whose second- and third-generation cameras on the SPT, SPTpol \citep{austermann12} and SPT-3G \citep{sobrin22} respectively, have been equipped with polarimeters and were used to observe the mm-wave sky in Stokes parameters I, Q, and U.
The significantly enhanced instantaneous polarization sensitivity of the SPT-3G camera allows us to monitor the polarization characteristics of numerous bright or highly polarized blazars with an almost daily cadence over periods of many years.

In this work, we use data from observations of the $\sim$1500~deg$^2$ SPT-3G Main field from March 2019 to November 2023 to examine the mm-wave EVPA variability of eight blazars. 
Specifically, we explore whether EVPA swings occur in the mm-wave, and, in combination with data from the Large Area Telescope (LAT) on the \textit{Fermi} Gamma-ray Space Observatory \citep{atwood09}, whether we observe EVPA swings coincident with gamma-ray flares.

In Sec.~\ref{Sec:Observations and data reduction} we summarize the mm-wave and gamma-ray observations.
In Sec.~\ref{Sec:Analysis} we describe our methods for source selection, data cuts, adjustments made to the mm-wave polarization data, the criteria for EVPA swings, random walk simulations, gamma-ray flare fitting, and probabilities of coincidental  association between gamma-ray flares and EVPA swings.
In Sec.~\ref{Sec:Results} we present the results of the EVPA swing, variability, random walk simulation, and coincidental gamma-ray flare association analysis.
In Sec.~\ref{Sec:Conclusion} we discuss and summarize the results. 

\section{Observations and data reduction}
\label{Sec:Observations and data reduction}
\subsection{mm-wave observations}
\label{SubSec: mm-wave observations}
The SPT is a 10-meter telescope located at the Amundsen-Scott South Pole Station in Antarctica. 
The SPT is designed to measure the CMB by repeatedly observing large patches of the sky at mm wavelengths.
The SPT-3G camera \citep{sobrin22}, installed in 2017, is sensitive to total intensity and linear polarization in three bands centered at roughly 95, 150, and 220~GHz (3.2, 2.0, and 1.4~mm) with angular resolution of $\sim$1$^\prime$.
The SPT-3G Main field is a $\sim$1500~deg$^2$ region of the southern sky, with the nominal full-coverage region extending from $-42^\circ$ to $-70^\circ$ in declination and  $20^\text{h}40^\text{m}0^\text{s}$ to $3^\text{h}20^\text{m}0^\text{s}$ in right ascension.

The SPT-3G Main field is composed of four sub-fields, with each sub-field covering the entire right ascension range and one-fourth of the declination range. 
Each sub-field is observed by the SPT in a raster pattern over the course of two hours.
The Main field is generally monitored by the SPT throughout the year, except for a gap during the austral summer for sun avoidance, with a semi-regular cadence ranging from two hours to one and a half days.

The data processing in this work, including cuts, calibration, filtering of time-ordered data, and map making closely follows the procedure described in \citet{quan26} and \citet{archipley26}. We briefly outline the standard procedure and discuss operations specific to this work. In the standard procedure, data from any detector not passing a standard set of tests (including stability and response to an internal calibration source) are cut. The remaining data are then converted from digital counts to equivalent CMB temperature using the combination of response to a dedicated observation of a Galactic HII region and to the internal calibration source. The data are then filtered to suppress low-frequency noise (particularly from the atmosphere) and to reduce aliasing when the data are binned into pixels.

In standard analyses, the detector time-ordered data from a single observations are then binned into pixels on a grid spanning the full $\sim$1500~deg$^2$ Main field.
For this work, we create 30$^\prime\times$30$^\prime$ maps, referred to as thumbnails, from the detector time-ordered data around the 168 point sources within the nominal boundaries of the SPT-3G Main field that have a 150~GHz flux density $>50$ mJy.

A matched filter \citep[e.g.,][]{zubeldia21} is applied to the Stokes I, Q, and U thumbnails at each frequency for every observation.
Within these thumbnails, each pixel has a 3$\times$3 weight matrix, which is proportional to the expected inverse covariance of the values of the three Stokes parameters in the pixel.
The matched filter implicitly assumes uniform noise, and, to avoid instances where this assumption is not true, we cut thumbnails based on the ratio of maximum and minimum values for the total intensity component of the weight matrix ($W$) for each pixel (keeping observations for which $\mathrm{max}(W)/\mathrm{min}(W) < 10$).
If this weight matrix is poorly conditioned, it can result in spuriously large Q and U pixel values.
To improve the conditioning of linear polarization thumbnails, Q and U thumbnail pixels are down-sampled by a factor of two. 
After down-sampling, thumbnails that are still poorly conditioned (maximum condition number $>10$) are cut.
The cuts on the min-max weight ratio and poorly conditioned thumbnails result in a loss of 23\% of all thumbnails, primarily those located near the sub-field boundaries.

As described in \citet{quan26}, the absolute calibration of the Main field data comes from comparing a Stokes I map coadded over many observations to a map of the same part of the sky from the \textit{Planck} satellite \citep{planck15-9}. We apply this (per-band) calibration factor to every thumbnail map. As a data quality check, we calculate absolute calibration values for every individual observation using the same technique, and we inspect the distribution of single-observation calibration values.
Observations with outlier calibration values ($>5\sigma$) are cut, resulting in a loss of 5.93\% of thumbnails.

The uncertainty on the I, Q, and U flux density values for each per-observation thumbnail is calculated by measuring the median absolute deviation of $10^{\prime}\times30^{\prime}$ slices of the top and bottom of the filtered thumbnail in order to avoid filtering wings from the source. 

For weakly polarized sources, it is crucial to correct any systematic effects that can cause signal to ``leak'' from the I thumbnails to the Q and U thumbnails. Various instrumental effects can cause I-to-Q/U leakage; for a differential polarimeter such as SPT-3G, in which Q and U are estimated by effectively differencing the signal from two orthogonally oriented detectors in a focal plane pixel, the most important mechanisms are mismatches in gain and beam properties between detectors in a pixel pair \citep{hu03}. We expect beam properties to be constant across the full Main field, but the gain-related leakage will depend on which HII region is used as the calibration source (for details, see \citealt{quan26}.)
We thus calculate the I-to-Q and I-to-U leakage coefficients separately in two halves of the Main field, split by declination (because observations of the two northernmost fields use RCW38 as the calibrator, and the two southernmost fields use Mat5a, \citealt{quan26}). We assume that the total EVPA distribution across all sources is random and uniform, and thus the average Q/I and U/I values should be zero.
We assume the leakage is not time-dependent and compute the leakage coefficients by averaging Q/I and U/I across the sources in each declination range (86 sources at $\delta > -56^\circ$, 82 sources at $\delta < -56^\circ$) over the entire five-year period, and adjusting such that they are equal to zero.
The Q and U leakage coefficients are $-0.237$\% and $-0.466$\% respectively, at 95~GHz, and $-0.285$\% and $-1.34$\% respectively, at 150~GHz for $\delta > -56^\circ$, and $-0.285$\% and $-0.168$\% respectively, at 95~GHz, and $-0.128$\% and $-0.653$\% respectively, at 150~GHz for $\delta < -56^\circ$.\footnote{We note that these are not consistent with the (purely gain-related) coefficients estimated in \citet{quan26}; this is because the analysis here is also sensitive to the higher-order leakage terms, and the single coefficient estimated here is an average of the leakage over a range of angular scales set by the matched filter.}

Observations are subsequently binned and averaged to improve the signal-to-noise ratio (SNR) of polarization measurements, with approximately two-day bins calculated according to the schedule of the SPT's cryogenic refrigerator.
This binning results in light curves that have an average cadence of two days, however, due to cut observations, their time differences vary.

The per-observation Q and U noise levels of the 220~GHz band, with an average of 58.6 mJy, is significantly higher than the Q and U noise levels of the 95 and 150~GHz bands, with averages of 12.8 and 14.2 mJy respectively.
For this reason the 220~GHz frequency band is excluded from the polarization analysis.

\subsection{Gamma-ray observations}
\label{SubSec:Gamma-ray observations}
The gamma-ray data utilized in this study are obtained from the LAT aboard the \textit{Fermi} Gamma-ray Space Telescope. We use the version of LAT data, in an energy band ranging from 0.1-100 GeV, processed and published by the \textit{Fermi} LAT Light Curve Repository (LCR, \citealt{abdollahi23}).
In this study, we use a three-day cadence with a 3$\sigma$ detection threshold, limited to the observation window of the SPT-3G Main field five-year survey (March 2019 to November 2023).

\subsection{ACT Data}
\label{SubSec:ACT Data}
One source in the sample, SPT-S~J021045-5101.0, has contemporaneous mm-wave data from ACT. 
ACT was a six-meter telescope dedicated to measuring the intensity and linear polarization anisotropy of the CMB. 
From 2017 to 2022, the telescope was equipped with the Advanced ACTPol camera, which featured five bands from 28 to 230~GHz \citep{henderson16}.
In part because of its location in the Atacama Desert in Chile (latitude $-22^{\circ}57' 31''$), the ACT survey strategy was able to cover a much larger area of the sky than SPT, with the final ACT DR6 data set covering nearly 50\% of the sky \citep{naess25}.
Due to the larger survey area, the ACT data on SPT-S~J021045-5101.0 are more widely spaced in time and shallower than those from SPT-3G, but they serve as a useful independent cross-check for SPT measurements of EVPA.
The average per-observation 90 and 150~GHz Q and U noise levels of ACT data are 25.3 and 32.6 mJy respectively, roughly twice the SPT-3G average 95 and 150~GHz Q and U noise levels of 12.8 and 14.2 mJy respectively.

\begin{figure*}[htbp] 
    \plotone{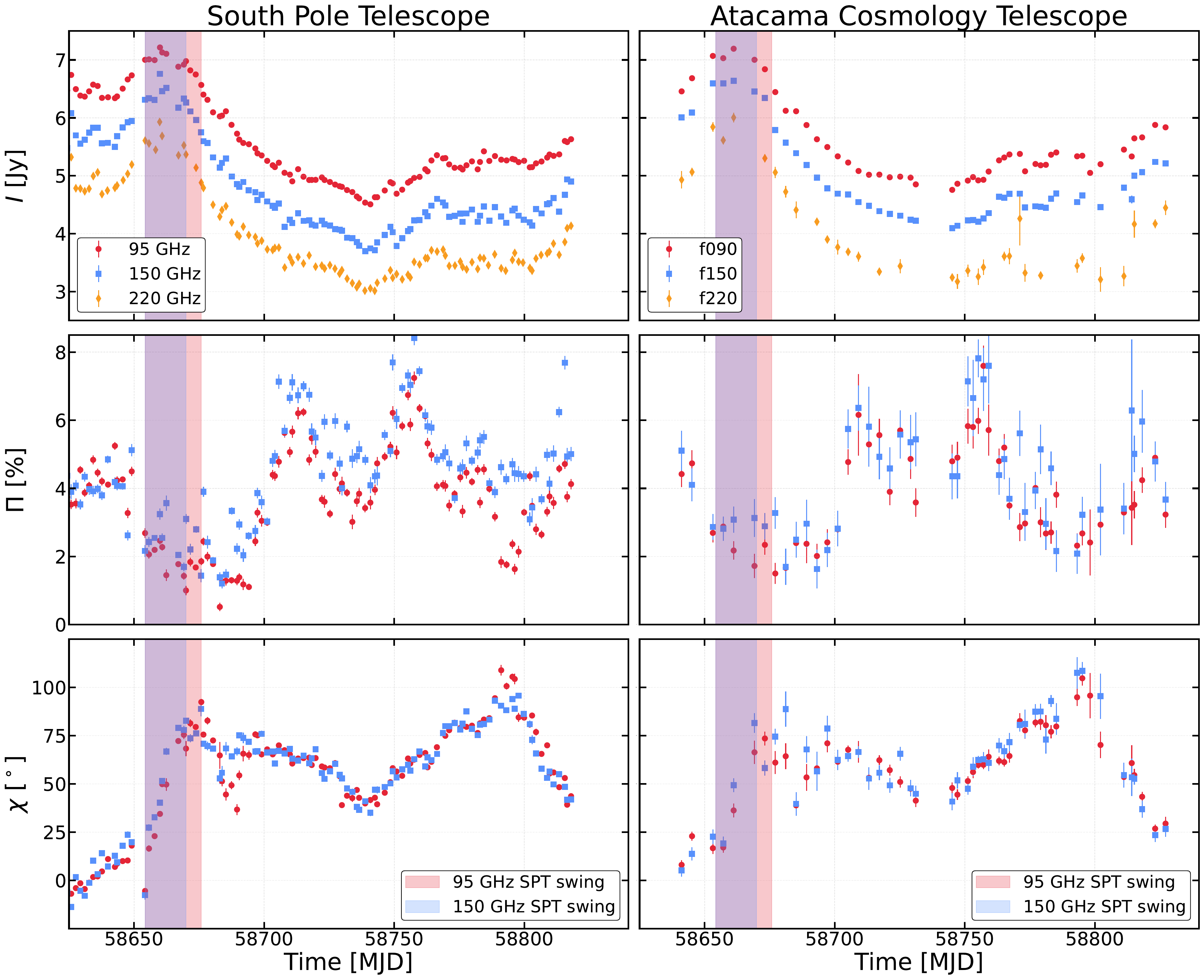}
    \caption{    
    Six-month light curves of SPT-S~J021045-5101.0 from SPT data binned in two-day intervals (left) and ACT data with typical separation of four days during this time interval (right).
    In each panel, from top to bottom, the rows are as follows: (1) Stokes I flux density in Jy for the 95 (red circle), 150 (blue square), and 220 (orange diamond) GHz frequency bands, (2) PD ($\Pi$), (3) adjusted EVPA ($\chi$) and identified SPT-3G EVPA swings in the 95 (red highlight) and 150~GHz band (blue highlight).
    This six-month period was chosen due to the high SNR of this source during this interval and its overlap between SPT and ACT data.
    This is an example of a mm-wave light curve as well as the typical result of the EVPA swing identifier described in Sec.~\ref{section:identify_swings}.
    While the EVPA swing identified by SPT-3G is not formally detected in the ACT data, the ACT data shows the same qualitative trend as the SPT-3G data over the same period.
    This provides an independent qualitative confirmation of the SPT-3G data, and serves as an example of the importance of cadence when detecting these EVPA swings.
    }
    \label{fig:SPTxACT}
\end{figure*}

The ACT light curve for SPT-S~J021045-5101.0 was obtained through private communication, with data cuts and calibration details explained in a forthcoming paper (Ma et al, in prep.).
A sample of this light curve is shown in Fig.~\ref{fig:SPTxACT} over the same time range as the SPT-3G observations.
Each ACT light curve data point is calculated by making dedicated source thumbnail maps using data combining a single night of observations.
Gaps in the light curve may be due to weather events or other observatory maintenance and down-time.

The ACT light curve for SPT-S~J021045-5101.0 provides an independent cross check on SPT-3G polarization measurements.
The six-month period shown in Fig.~\ref{fig:SPTxACT} was chosen for the high SNR of the source during this interval and the overlap between SPT and ACT data.
ACT typically observes this particular source with a $\sim$daily cadence, but over this particular time interval the median cadence is four days.
As a result, the EVPA swing identified in the SPT-3G data does not meet the identification requirements described in Sec.~\ref{section:identify_swings} when applied to the ACT light curve, despite the ACT total intensity, PD, and EVPA measurements following the same qualitative trends as those from SPT-3G over the same time period. 
This agreement also provides qualitative confirmation that the SPT-3G I, Q, and U flux density measurements are not significantly impacted by systematic error.

\section{Analysis}
\label{Sec:Analysis}
Our analysis of EVPA properties, the algorithm to identify EVPA swings, and our treatment of the $180^\circ$ ambiguity follow methods similar to those found in \citet{blinov15}, and we outline those steps here.
\subsection{Polarization parameters and corrections}
Following the measurement of the I, Q, and U flux density values, we calculate the initial values for the PD ($\Pi_0$) and EVPA ($\chi_0$) as below:
\begin{eqnarray}
    \label{eq1}
        \Pi_0 & = & \frac{\sqrt{Q^2 + U^2}}{I} \\
        \chi_0 & = &  \frac{1}{2}\text{arctan}\left(\frac{U}{Q} \right)\,.
\end{eqnarray}

Prior to performing the variability analysis on the polarization features, it is necessary to make corrections to the PD and the error associated with the EVPA.
Since the PD is a the result of squared Q and U values, any noise in the observation will produce a bias which raises the resulting PD value.
This bias is corrected by using the method described in \citet{plaszczynski14}:
\begin{eqnarray}
    \label{eq3}
    b^2&=&\frac{q^2\sigma^2_{u} + u^2 \sigma_q^2}{q^2+u^2} \\
    \Pi&=&\Pi_{0} - b^2 \frac{1 - e^{-\Pi_{0}^2/b^2}}{2\Pi_{0}}\,,
\end{eqnarray}
where $q=Q/I$ and $u = U/I$.

To compute the EVPA error, we employ the method outlined in \citet{naghizadeh-khouei93}, which takes into account the fact that the EVPA error is non-Gaussian and proportional to the PD SNR.
The probability distribution of observed EVPA $\chi$ given the true EVPA $\chi^\prime$ and PD SNR $\text{SNR}_\Pi^\prime \equiv \Pi^\prime/\sigma_\Pi$ is expressed as:

\begin{multline}
\label{EVPAprob}
G(\chi;\chi^\prime, \text{SNR}_\Pi^\prime) =  \\
\frac{1}{\sqrt{\pi}} \left\{\frac{1}{\sqrt{\pi}} + \eta e^{\eta^2}[1 + \erf(\eta)] \right\} e^{-\left( \frac{(\text{SNR}_\Pi^\prime)^2}{2}\right)}
\end{multline}
where $\eta = \frac{\text{SNR}_\Pi^\prime}{\sqrt{2}} \cos{2(\chi - \chi^\prime)}$, and $\erf$ is the Gaussian error function. To specifically extract the EVPA uncertainty (without regard for the mean value), we set $\chi^\prime=0$ and solve for the value of $\sigma_\chi$ that minimizes the difference between the integrated probability distribution and the target probability interval ($\pm 1\sigma$ in this case): 
\begin{eqnarray}
    \sigma_\chi = \argmin_x \left ( \label{EVPAerr}
    \int_{-x}^x G(\chi; 0, \text{SNR}_\Pi^\prime) d\chi - 0.6826 \right ).
\end{eqnarray}
The integral is evaluated numerically for each data point with $\left( \text{SNR}_\Pi \right)<10$.
For data points with $\left( \text{SNR}_\Pi \right)>10$ the EVPA probability distribution is well-approximated by a Gaussian, and we instead use the analytic approximation:
\begin{eqnarray}
    \label{EVPAerrapprox}
    \sigma_\chi = 28.65^\circ \frac{1}{\text{SNR}_\Pi}.
\end{eqnarray}

\subsection{Polarization source selection and data cuts}
\label{subsec:Polarization source selection}
In order to both confidently measure EVPA and identify day-time-scale EVPA swings, we limit our study to sources with enough polarized flux to produce a low-error EVPA measurement in a single two-day coadd.
We begin with the sample of 168 sources described in Sec.~\ref{SubSec: mm-wave observations}. We then compute the average PD SNR ($\text{SNR}_{\Pi,\nu}$, computed with the bias-corrected PD) for both 95 and 150~GHz, for each observation over the entire five-year observing period.
All sources with $\langle\text{SNR}_{\Pi,95}\rangle < 3$ or $\langle\text{SNR}_{\Pi,150}\rangle < 3$ are cut from our sample.
This cut results in a loss of 160 sources, leaving eight.

Table~\ref{table:main_sample} shows the names of these eight sources in the SPT-3G and Parkes (PKS, \citealt{wright90}) catalogs, the redshift of each source, the average PD SNR in each band, and the amount of data cut from each source's light curve in each band using the per-observation cuts detailed below. 
The combination of the fact that all of these sources are in the PKS radio catalog, the Roma-BZCAT \citep{massaro15a}, and the bright end of the SPT-3G mm-wave catalog, and the fact that they are all at high redshift, indicates that they are almost certainly all blazars.
We will subsequently refer to this sample as our ``Main Sample of Blazars.''
To examine a potential correlation between gamma-ray flares and EVPA swings, we create an additional sub-sample from the Main Sample.
The new gamma-ray sample is composed of four blazars (seen in Table~\ref{table:main_sample}), each having at least ten consecutive $3\sigma$ gamma-ray detections, with no more than seven days between adjacent detections, at some point during the five-year duration of the SPT-3G survey.

\begin{deluxetable*}{lccccccc}
\digitalasset
\tablewidth{0pt}
\tablecaption{\label{table:main_sample} Main Sample of Blazars}
\tablehead{
\colhead{SPT ID} & \colhead{PKS ID} & \colhead{z} & \colhead{$\langle \text{SNR}_{\Pi,95} \rangle$} & \colhead{$\langle \text{SNR}_{\Pi,150} \rangle$} & \colhead{$\%\chi_{95} \text{ cut}$} & \colhead{$\%\chi_{150} \text{ cut}$} 
}
\startdata
SPT-S J021045-5101.0* &  PKS 0208-512 & 1.003 & 15.2 & 12.7 & 13.0\% & 17.3\% \\ 
SPT-S J232918-4730.3 & PKS 2326-477 & 1.306 & 4.67 & 3.21 & 22.7\% & 48.5\% \\
SPT-S J205616-4714.8* & PKS 2052-47 & 1.489 & 3.39 & 3.15 & 62.7\% & 65.5\% \\
SPT-S J030956-6058.6* & PKS 0308-611 & 1.480 & 4.11 & 3.27 & 38.9\% & 53.6\% \\ 
SPT-S J013305-5200.1* & PKS 0131-522 & 0.925 & 3.99 & 3.26 & 36.5\% & 47.8\% \\  
SPT-S J005846-5659.1 & PKS 0056-572 & 1.460 & 4.02 & 3.60 & 58.5\% & 62.6\% \\ 
SPT-S J025329-5441.8 & PKS 0252-549 & 0.537 & 6.96 & 5.40 & 7.20\% & 17.2\% \\
SPT-S J231545-5018.6 & PKS 2312-505 & 0.811 & 6.94 & 5.10 & 15.8\% & 28.0\% \\ \hline
\enddata
\tablecomments{SPT-3G blazars with $\langle \text{SNR}_{\Pi,95} \rangle > 3$ and $\langle \text{SNR}_{\Pi,150} \rangle > 3$. Table columns are as follows: (1,2) SPT and Parkes Catalog (PKS, \citealt{wright90}) identifiers where $^{*}$ indicates that the blazar is part of the gamma-ray sample; (3) redshift measurement \citep{veron-cetty10, chen24}; (4,5) Average PD SNR for the 95 and 150~GHz bands; (6,7) percentage of data cut from the 95 and 150~GHz bands.}
\end{deluxetable*}

We also cut data from our blazar light curves at the per-observation level. 
Below a PD SNR of three, the EVPA becomes highly uncertain \citep{angelakis16}.
These highly uncertain EVPA data points result in accidental triggering of the $180^\circ$ adjustment described in Sec.~\ref{SubSec:180 Ambiguity}, causing the EVPA light curve to jump around wildly.
In order to reduce this accidental triggering of the $180^\circ$ adjustment, data points in the $\chi$ light curve with a $\text{SNR}_{\Pi,\nu}< 3$ are cut.
The percentage of EVPA observations removed by this cut (\%$\chi_\nu$ cut) in both 95 and 150~GHz is shown in Table~\ref{table:main_sample}.

\subsection{180° Ambiguity}
\label{SubSec:180 Ambiguity}
EVPA is defined within a 180$^\circ$ interval; in this work we use the interval $-90^\circ \leq \chi_0 < 90^\circ$.
When points reach the boundary of this interval, they are wrapped around to the opposite side of the interval (i.e. a 2$^\circ$ increase in an EVPA of 89$^\circ$ will result in an EVPA of $-89^\circ$).
This presents a challenge for measuring EVPA variability and identifying EVPA swings, since both the amplitude and direction of $\Delta \chi$ between consecutive points remain ambiguous.

In order to resolve this issue, we assume the minimum distance between consecutive data points, and $\chi_0$ is adjusted in 180$^\circ$ intervals to achieve this minimum distance:
\begin{equation} \label{eq:180adj_app}
    \chi = \chi_0 - k\cdot 180^\circ 
\end{equation}
where
\begin{equation} \label{eq:180adj}
    k = \text{round}\left(\frac{\chi_{0,i} - \chi_{i-1}}{180}\right) \, .
\end{equation}
From equation~\ref{eq:180adj_app} and \ref{eq:180adj} it is clear that EVPA observations differing by integer multiples of 180$^\circ$ (e.g., 90$^\circ$ and 450$^\circ$) are equivalent measurements. 
It is also important to note that this adjustment is applied to all adjacent data points within the same observing season, but not across seasonal gaps.

\subsection{Probability of correct measurement}
\label{SubSec:Probability of correct measurement}

The key assumption in our treatment of the 180$^\circ$ ambiguity is that the true difference between two adjacent EVPA observations is the minimum one.
However, as the rotation rate increases and the difference between adjacent data points approaches $90^\circ$, this assumption of minimum difference becomes less likely to be true.
To assess the probability that the change between two adjacent EVPA data points is measured correctly, we utilize the methodology outlined in~\citet{kiehlmann21} and summarized here.

For every measurement of EVPA ($\chi_i$,$t_i$), the time differences $\Delta t_{i,j}$ and change in EVPA $|\Delta \chi_{i,j}|$ are computed for all points $t_j>t_i$ for each of the blazars presented in Table~\ref{table:main_sample}.
This method will also be used to investigate the variability of each individual source in Sec.~\ref{SubSec:Rotators vs non-rotators}.

Each $|\Delta \chi_{i,j}|$ value is then binned according to their associated $\Delta t_{i,j}$ into the following bins: $\Delta t_{i,j} = 1.5\pm1.5, 4.5\pm1.5$, and $7.5\pm1.5$ days.
The EVPA variability is modeled as a log-normal distribution:
\begin{equation}
\mathcal{LN}_{\text{model}}(x;\mu, \sigma) = \lim_{N\xrightarrow{}\infty}\sum_{n = 0}^N \mathcal{LN}(x_n;\mu,\sigma),
\end{equation}
where
\begin{equation}
x_n = 90[n+m(n)] + (-1)^n x,
\end{equation}
and
\begin{equation}
\mathcal{LN}(x;\mu, \sigma) = \frac{1}{\sqrt{2\pi}\sigma x}\exp{\left( -\frac{(\ln{x}-\mu)^2}{2\sigma^2}\right)}.
\end{equation}
This model is then fit to the distribution of $|\Delta \chi_{i,j}|$ for each bin to calculate best-fit values of ($\mu(\Delta t)$, $\sigma(\Delta t)$).
We find values of ($1.98$, $0.99$), ($2.05$, $1.04$), and ($2.16$, $1.04$) for the $\Delta t_{i,j} = 1.5\pm1.5, 4.5\pm1.5$, and $7.5\pm1.5$ day bins, respectively.
The probability that any given $\Delta \chi_{i,j}$ is measured correctly is:

\begin{equation}
P(\Delta \chi_{i,j}, \Delta t) = \frac{\mathcal{LN}( \Delta\chi_{i,j}; \mu (\Delta t), \sigma (\Delta t))}{\mathcal{LN}_{\text{model}}(\Delta\chi_{i,j}; \mu(\Delta t), \sigma(\Delta t))}.
\end{equation}
The probability of correct measurement is then calculated for all neighboring data points.

\subsection{Identifying EVPA swings}\label{section:identify_swings}

In order to identify EVPA swings, the EVPA light curve must be segmented to define the start and end of these large changes in EVPA.
Segments of the EVPA light curve are created using the methodology defined in~\citet{kiehlmann16}, where start and end points for segments are determined by significant changes in the sign of the EVPA time derivative or where the time difference between consecutive points is larger than seven days ($\Delta t_{i,i+1} > 7$ days).
Significant changes in the sign of the time derivative occur when both the sign changes and
\begin{equation}
\Delta \chi_{i,i+1} > 3\sqrt{\sigma_{\chi,i}^2 + \sigma_{\chi,i+1}^2}.
\end{equation}
Each segment created is characterized using six parameters: number of data points in the segment ($N$), number of significant changes in the segment ($N_{\text{sig}}$), difference in EVPA between starting and ending data points ($\Delta\chi$), duration of the segment ($\Delta t$), average rate of change ($\langle\dot{\chi}\rangle$), and the probability of correct measurement ($P_{\text{swing}}$) for the segment which is the product of probabilities in the segment.
In this work, an EVPA swing is defined as any segment with $P_{\text{swing}}>68\%$, $N\geq 3$, $N_{\text{sig}} \geq 2$, and $|\Delta \chi| > 90^\circ$.
Although there is no universal definition of an EVPA swing, this definition is similar to those used in previous studies \citep[e.g.,][]{blinov15}.

\subsection{Random walk simulations}
\label{SubSec:Random walk simulations analysis}
A possible physical mechanism that could produce EVPA swings is a stochastic process originating from turbulence in the magnetic field of the jet \citep{moore82}.
In a model of this process, the blazar jet is composed of numerous small independent cells, each possessing the same PD but uncorrelated polarization angles, which vary over time (also in an uncorrelated way). At any snapshot in time, for an observation that does not resolve the jet structure, the observed polarization angle will be the average of the angles of the individual cells, and this average will fluctuate in a way that can produce apparent EVPA swings.
To determine if the observed variability in EVPA can be reproduced by this model of stochastic fluctuations in polarization parameters, we perform random walk simulations similar to those described in \citet{kiehlmann16}.

We generate 1000 random walk simulations of the 95 and 150~GHz bands for each blazar in the Main Sample.
The time steps between points are drawn from the CDF of observed time differences between observations until the total simulated time matches the total observed time.
The PD of every cell is set as the expected PD of synchrotron radiation in a uniform magnetic field, $\Pi_{\text{exp}} = (p + 1)/(p+ 7/3)\approx72\%$ \citep[e.g.,][]{longair11}, where $p$ is the slope of the electron energy spectrum.
The number of cells needed to replicate the observed degree of polarization ($\Pi_{\text{obs}}$) with individual cells exhibiting random polarization vector orientations is found in \citet{smith12b} to be 
\begin{equation}
 N_{\text{cells}} = \left( \frac{\Pi_{\text{exp}}}{ \langle \Pi_{\text{obs}}\rangle} \right)^2.
\end{equation}

At each time step, a specific number of cells are randomly selected, and the polarization vector direction of those cells is changed randomly.
To replicate the observed variability in the PD, we randomly select
\begin{equation}
N_{\text{var}} = \frac{\sigma(\Pi_{\text{obs}})}{\langle \Pi_{\text{obs}} \rangle}N_{\text{cells}}
\end{equation}
cells to be changed each time step, where $\sigma(\Pi_{\text{obs}})$ is the standard deviation of the observed PD.
Observational noise, drawn from the CDF of observed noise, is then added to the simulated polarization values.
EVPA and PD light curves are constructed, then rotations and EVPA swings are calculated using the same method and requirements described in Sec.~\ref{section:identify_swings}.
Only simulations that are able to replicate $\Pi_{\text{obs}}$ and $\sigma(\Pi_{\text{obs}})$ within $20\%$ are considered.
Of the 16,000 total random walk simulations, $47.2\%$ fell within the $20\%$ threshold while the other $52.8\%$ of simulations fall outside of the threshold purely from stochastic fluctuations.
In Fig.~\ref{fig:Swing_distributions} and \ref{fig:N_swings}, the simulations for the 95 and 150~GHz bands are combined due to their similarity.

\subsection{Fitting gamma-ray flares}
\label{SubSec:Fitting gamma-ray flares}
To identify gamma-ray flares in the sub-sample of our blazars with robust Fermi-LAT data (see Sec.~\ref{subsec:Polarization source selection}), the gamma-ray light curves for these sources are fit using a standard flare profile characterized by an exponential rise and decay \citep[e.g.,][]{chatterjee12}:
\begin{equation}
F(t) = F_c + \sum_{i = 1}^{N_{\text{flare}}} F_{p,i}\left( e^{\frac{t_{p,i} - t}{T_{r,i}}} + e^{\frac{t - t_{p,i}}{T_{d,i}}} \right),
\end{equation}
where $F_c$ is a constant flux, $i$ represents an individual flare, $F_{p,i}$ is the flux value for the flare amplitude, $t_{p,i}$ is the flare time, $T_{r,i}$ is the rise time, and $T_{d,i}$ is the decay time.
To avoid overfitting, boundaries are placed on the fit parameters such that the distance between two flares is never smaller than three days and the flux of the flare is never greater than two times the maximum observed flux of \textit{Fermi} detections within the SPT time frame.
This profile was fit to each continuous segment of three-day $3\sigma$ \textit{Fermi} detections, with integer values of $N_{\text{flare}} = 0,1,2,\dots,  N_{\gamma, \text{det}}/3$, where $N_{\gamma,\text{det}}$ is the number of \textit{Fermi} detections.
For each $N_{\text{flare}}$ fit, the Akaike Information Criterion (AIC) is computed to penalize models with additional flares for their increased number of free parameters, preventing overfitting by selecting the simplest model that adequately describes the data.
The $N_{\text{flare}}$ model with the lowest AIC value is chosen as the optimal fit.

The time lag between the identified EVPA swing and its closest gamma-ray flare is calculated as the time difference between the start of the EVPA swing and start of the flare, $\tau_{\text{obs}} = t_{\text{swing}, \text{start}} - (t_{p,i} - T_{r,i})$.
In some cases, multiple EVPA swings are associated with the same flare; in these instances, we only consider the EVPA swings with the minimum $|\tau_{\text{obs}}|$, and the 95 and 150~GHz bands are treated independently.

\subsection{Probability of coincidental flare associations}
\label{SubSec:Probability of coincidental flare associations}

In order to assess the probability of coincidental association between EVPA swings and gamma-ray flares, for every source at both frequencies, we uniformly randomize the time of each identified EVPA swing bounded by the first and last observation time.
The distance to the flares for the randomly positioned EVPA swings is then computed utilizing the same method as the observed data.
The probability of a coincidental association, $P(\tau)$, is then calculated as the fraction of simulations for that source and that frequency in which $\text{min}(|\tau_{\text{sim}}|) < |\tau_{\text{obs}}|$.
Similarly, the probability of a coincidental time lag with a flare of equal height or greater, $P(\tau, F_{\text{peak}})$, is calculated as the number of simulations where $\text{min}(|\tau_{\text{sim}}|) < |\tau_{\text{obs}}|$ and $F_{\text{peak}, \text{sim}} \geq F_{\text{peak}, \text{obs}}$.

\section{Results}
\label{Sec:Results}
\subsection{Observed EVPA swings}
\label{SubSec: Observed Characteristics of Polarization Angle Swings}

In the five years of observations of the eight blazars in our Main Sample, we identify a total of 14 EVPA swings.
Of the 14 total EVPA swings, six are identified in the 95~GHz band, and eight are identified in the 150~GHz band.
All EVPA swings are identified in the light curves of sources in our gamma-ray subsample, and all sources in our gamma-ray subsample have at least one EVPA swing. 
Assuming no connection between gamma-ray emission and EVPA swings, the probability of all swings occurring exclusively in gamma-ray sources by chance is 0.78\%. We investigate the connection between EVPA swings and gamma-ray activity further in Sec.~\ref{SubSec:Observed time lags between flares and swings}.
Fig.~\ref{fig:SPTxACT} shows an example six-month SPT light curve with matching ACT data, with the EVPA swings detected by SPT-3G highlighted.
Full light curves for all blazars in the Main Sample containing flux density, PD, EVPA, and gamma-ray data are shown in Fig.~\ref{fig:gamma-ray_LCs} and \ref{fig:non_gamma-ray_LCs}, with observed EVPA swings in 95 and 150~GHz highlighted in red and blue, respectively.

\begin{figure*}[th]

    \plotone{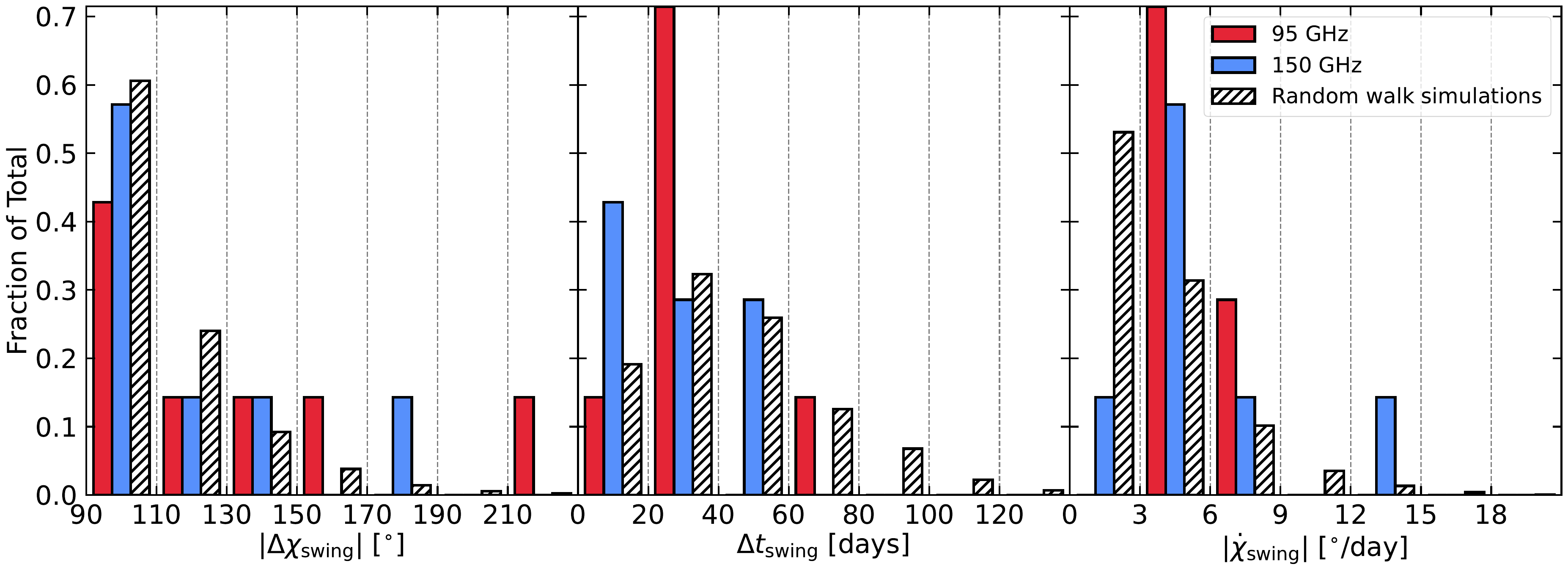}
    \caption{Distributions of Main Sample EVPA swing amplitudes (left), durations (middle), and rates (right), in the 95~GHz (red), 150~GHz (blue), and combined 95 and 150~GHz random walk simulations (hatched black).
    }
    \label{fig:Swing_distributions}
\end{figure*}

The distribution of the EVPA swing characteristics $|\Delta \chi_\text{swing}|$, $\Delta t_\text{swing}$, and $\langle \dot{\chi}_{\text{swing}}\rangle$ for both bands, and for the random walk simulations described in Sec.~\ref{SubSec:Random walk simulations analysis}, can be seen in Fig.~\ref{fig:Swing_distributions}. 
Two-sample Anderson-Darling (A-D) tests \citep[e.g.,][]{ivezic20} are performed on various EVPA swing and rotation parameter distributions, with the null hypothesis for all tests being that they are drawn from the same distribution.
All reported p-values are results of either these A-D tests, and we consider any test with a resulting p-value $<0.05$ as construing significant evidence that the two sets of points under test do not originate from the same underlying distribution.

As can be seen in Fig.~\ref{fig:Swing_distributions}, the observed EVPA swing amplitudes range from $90.1^\circ$ to $212^\circ$, with larger-amplitude swings being less common than smaller-amplitude swings in both bands.
The EVPA swing durations range from 11.0 to 60.2 days, with 12 of the 14 swings lasting less than 25 days.
The rotation rates of the observed EVPA swings range from $2.99^\circ/$day to $12.6^\circ/$day.
We find no significant difference between the 95 and 150~GHz bands in any of these swing characteristics with A-D p-values $> 0.25$ for all.
We note that our EVPA swing definition imposes minimum values on these quantities: amplitudes are limited to $|\Delta\chi|>90^\circ$ and durations to $\Delta t \gtrapprox 6$ days by the requirement of $N\geq 3$ data points at $\sim 2$-day cadence.
The maximum detectable rotation rate is additionally limited by the $\sim$2-day cadence, since faster rotations produce inter-point EVPA changes approaching $90^\circ$, reducing the probability of correct measurement below the $P_{\text{swing}} > 68\%$ threshold required for a valid swing.

\subsection{Comparison to random walk simulations}
\label{subsec:Random walk simulations results}

\begin{figure}[th]
    \plotone{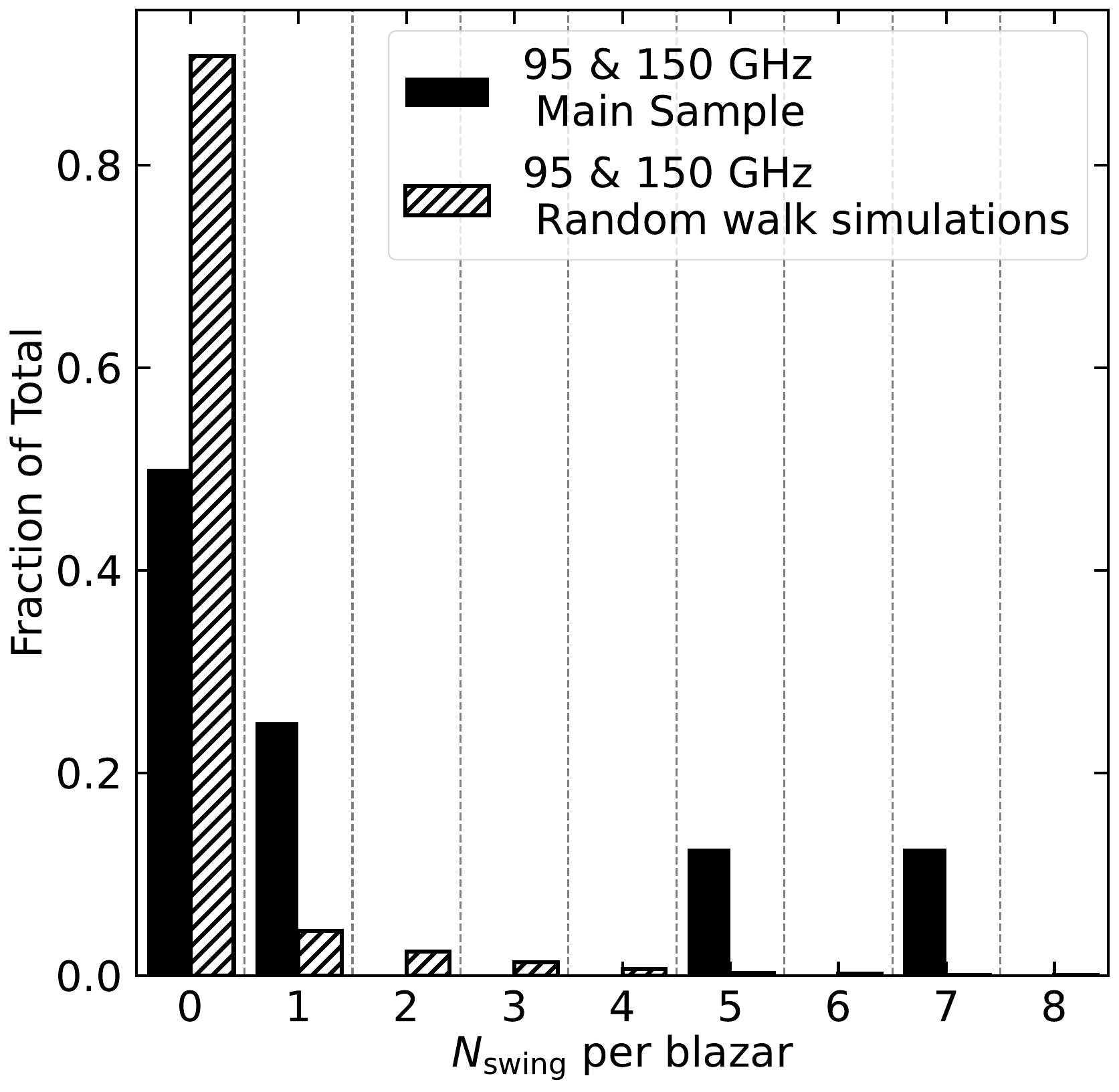}
    \caption{
    Distribution of the number of EVPA swings observed (in both 95 and 150~GHz) per blazar in the Main Sample (black) and in the random walk simulations (hatched black).
    The random walk simulations are unable to replicate the distribution of EVPA amplitudes, durations, and rotation rate, and the number of EVPA swings produced by each simulation is significantly lower (A-D p-value $<0.001$) than the sources in our observed data.
    }
    \label{fig:N_swings}
\end{figure}

When comparing observed distributions of EVPA swing characteristics to the random walk simulations, we combine the 95 and 150~GHz bands. 
We calculate the number of EVPA swings produced in each simulated light curve and compare to the observed number of EVPA swings in each observed blazar in Fig.~\ref{fig:N_swings}.
We find that the number of EVPA swings produced per simulated light curve is significantly smaller than those observed (A-D p-value $<0.001$).

The average EVPA swing amplitude produced by the simulations is $111^\circ$, which is smaller than the average observed EVPA swing amplitude of $124^\circ$, and the difference in these distributions is mildly significant (A-D p-value $= 0.035$).
The distributions of both the duration and the rotation rate differ even more significantly between random walk simulations and data. The average duration of the EVPA swings produced by the random walk simulations is $43.6$ days while the average duration of the observed EVPA swings is $26.0$ days, and the distributions are significantly different with an A-D p-value of $0.0066$. The random walk simulation EVPA swings also have a lower average rotation rate of $3.71$ $^\circ$/day compared to the average observed rotation rate of $5.60$ $^\circ$/day, and the distributions are found to be significantly different, with an A-D p-value $<0.001$.

\subsection{Variability analysis}
\label{SubSec:Rotators vs non-rotators}
Upon visual inspection of the EVPA light curves in Fig.~\ref{fig:gamma-ray_LCs} and \ref{fig:non_gamma-ray_LCs}, it is apparent that the EVPA variability differs significantly between sources.
For example, SPT-S~J205616-4714.8 has five EVPA swings and large two-day EVPA changes, while the five-year EVPA observations for SPT-S~J025329-5441.8 lie almost entirely within a $75^\circ$ interval. 
Following this observation, we assess the EVPA variability of each source, utilizing a method similar to that used to evaluate the probability of correct measurement in Sec.~\ref{SubSec:Probability of correct measurement}.

\begin{figure*}[th]
    \plotone{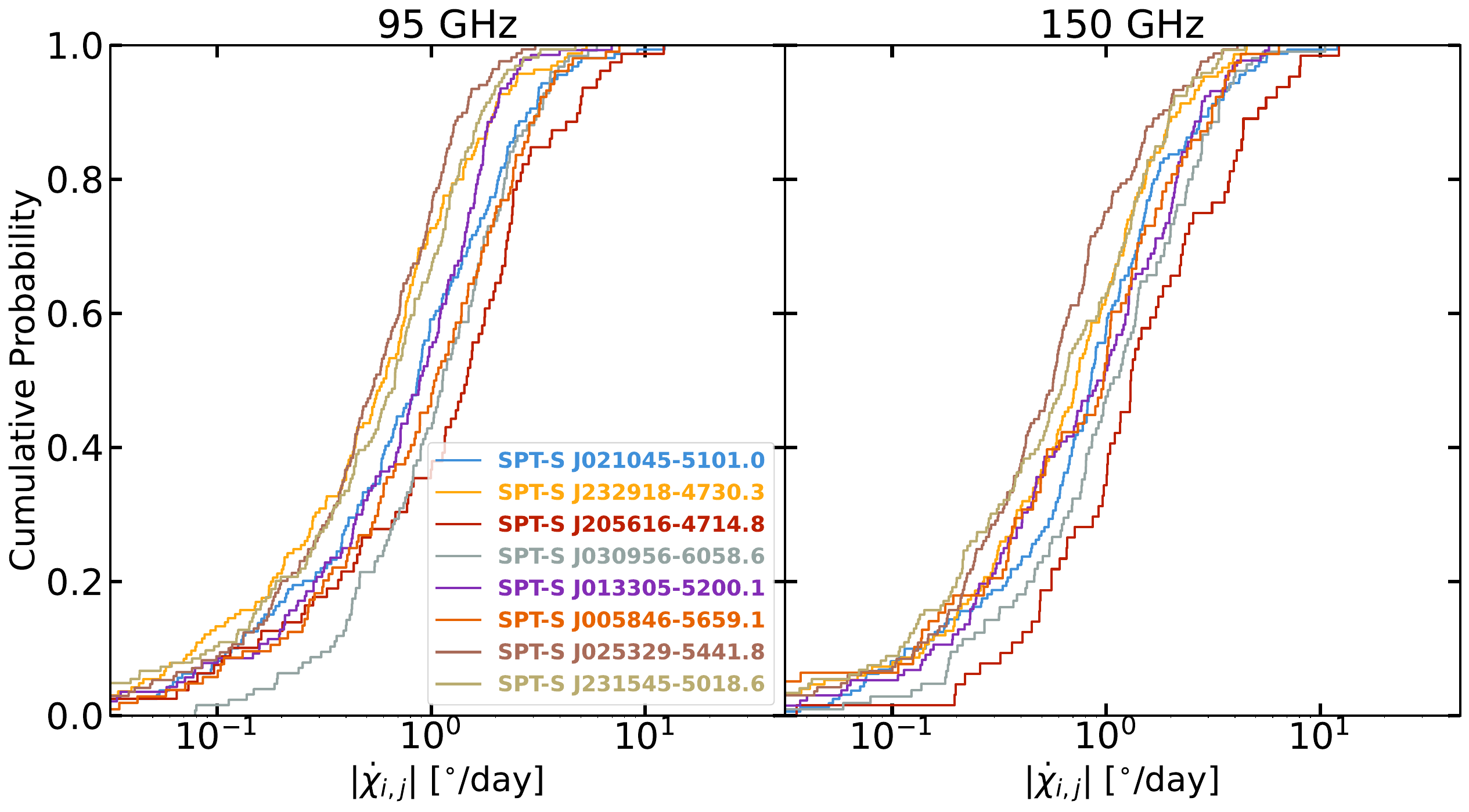}
    \caption{
    CDFs of the EVPA rotation rates for the eight blazars in the Main Sample at 95~GHz (left) and 150~GHz (right), computed from pairs with time differences of $4 < \Delta t_{i,j} < 10$ days on $\sim$7-day binned light curves. A distribution further to the left indicates lower variability while a distribution further to the right indicates higher variability. SPT-S~J205616-4714.8 (dark red) is a clear outlier, with its CDF sitting consistently to the right of all other sources at both frequencies, indicating systematically higher EVPA rotation rates on $\sim$7-day timescales. Three blazars, SPT-S~J232918-4730.3 (orange), SPT-S~J025329-5441.8 (brown) and SPT-S~J231545-5018.6 (tan), show lower variability than all other sources in the Main Sample in the 150~GHz band, and all sources except SPT-S~J232918-4730.3 in the 95~GHz band. These results are consistent with by-eye inspection of the EVPA light curves in Fig.~\ref{fig:gamma-ray_LCs} and \ref{fig:non_gamma-ray_LCs}.
    }
    \label{fig:evpa_affinity}
\end{figure*}

For the purpose of this variability analysis we further bin observations, according to the schedule of SPT's cryogenic refrigerator, into approximately seven-day bins to reduce the bias introduced by differences in per-observation EVPA uncertainties between sources. 
Then, for every measurement of EVPA ($\chi_i$, $t_i$), the time differences $\Delta t_{i,j}$ and EVPA rotation rate $|\Delta\chi_{i,j}/\Delta t_{i,j}|$ are computed for all points $t_j > t_i$ for each individual blazar.
Due to cut observations the time differences between adjacent binned points vary, and the separation between times assigned to adjacent bins can be smaller or larger than seven days.
We therefore take the distribution of rotation rates with $4<\Delta t_{i,j} < 10$ days, corresponding to pairs separated by approximately one bin width for each source in both 95 and 150~GHz.
The resulting rotation rate CDFs for each source for both 95 and 150~GHz with time differences of $4<\Delta t_{i,j}<10$ days can be seen in the top row of Fig.~\ref{fig:evpa_affinity}.

In the 150~GHz CDF, SPT-S~J205616-4714.8 is a clear outlier, with its CDF sitting consistently to the right of all other blazars, indicating systematically higher rotation rates on $\sim$7-day timescales.
On the other end, three blazars, SPT-S~J232918-4730.3, SPT-S~J025329-5441.8, and SPT-S~J231545-5018.6, show lower variability in both 95 and 150~GHz than all sources in the Main sample.
These results are consistent with by-eye inspection of the EVPA light curves in Figs.~\ref{fig:gamma-ray_LCs} and \ref{fig:non_gamma-ray_LCs}.

\subsection{Observed time lags between gamma-ray flares and EVPA swings}
\label{SubSec:Observed time lags between flares and swings}
\begin{figure*}[th]

    \plotone{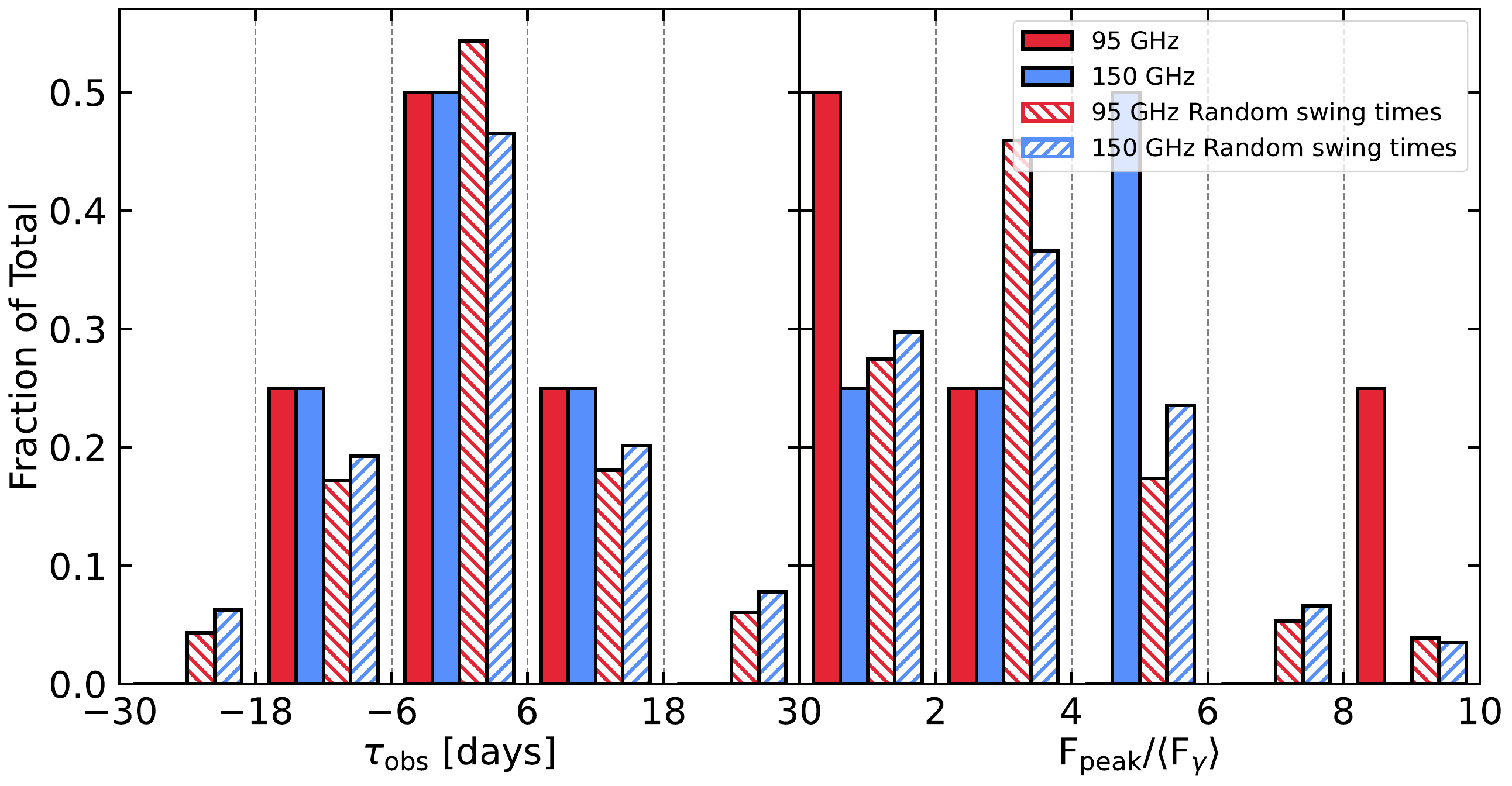}
    \caption{Distribution of time lags between EVPA swings and gamma-ray flares (left) and relative height of associated gamma-ray flares (right) for EVPA swings with $|\tau_{\text{obs}}|<30$ days for 95~GHz (red), 150~GHz (blue), 95~GHz simulations (hatched red), and 150~GHz simulations (hatched blue).
    Within this plot the dotted lines mark the bin edges of the histogram, all bins within these lines represent the same bin for the different data sets.
    While the gamma-ray associated EVPA swings show no significant preference for positive or negative time delays (A-D p-values $> 0.25$), six out of the eight EVPA swings have a positive time delay (see Table~\ref{table:gamma-ray_swings}). 
    These EVPA swings also show no significant preference for large or small gamma-ray flares.
    }
    \label{fig:tau}
\end{figure*}

Within the 14 EVPA swings observed in the 95 and 150~GHz bands, eight are identified with small time lags ($|\tau_{\text{obs}}| < 30$ days) to gamma-ray flares.
The distribution of observed time lags between these EVPA swings and the closest gamma-ray flares is illustrated in Fig.~\ref{fig:tau}, and the characteristics of each EVPA swing associated with a gamma-ray flare can be found in Table~\ref{table:gamma-ray_swings}.

\begin{deluxetable*}{ccccccccccc}
\digitalasset
\tablewidth{0pt}
\tablecaption{\label{table:gamma-ray_swings} EVPA swings associated with gamma-ray flares, with $|\tau_{\text{obs}}|<30$ days.}
\tablehead{
\colhead{SPT ID} & \colhead{$\nu$} & \colhead{Start} & \colhead{End} & \colhead{$\Delta \chi$} & \colhead{$\dot{\chi}$} & \colhead{$\tau_{\text{obs}}$} & \colhead{$F_{\text{peak}}$} & \colhead{P($\tau$)} & \colhead{P($\tau$,$F_{\text{peak}}$)} \\
\colhead{} & \colhead{GHz} & \colhead{MJD} & \colhead{MJD} & \colhead{$^\circ$} & \colhead{$^\circ$/day} & \colhead{days} & \colhead{$(\text{ph}\cdot 10^6)/(\text{cm}^2 \cdot \text{s})$} & \colhead{} & \colhead{} 
}
\startdata
SPT-S J021045-5101.0 & 95  & 58654 & 58675 & 98   &  4.53  &  2.84  & 0.912 & 0.0850 & 0.0533 \\
SPT-S J021045-5101.0 & 150 & 58654 & 58669 &  90   &  5.74  &  2.85  & 0.912 & 0.0920 & 0.0594 \\
SPT-S J205616-4714.8 & 95  & 58748 & 58808 & 212   &  3.52  &  0.149 & 2.23  & 0.0103 & 0.000614 \\
SPT-S J205616-4714.8 & 95  & 58952 & 58982 & 144   &  4.82  &  -11.1 & 0.43  & 0.476 & 0.406 \\
SPT-S J205616-4714.8 & 95  & 59042 & 59062 & $-$102  & $-$5.08  & 12.1   & 0.145 & 0.489  & 0.447 \\
SPT-S J205616-4714.8 & 150 & 59040 & 59066 & $-$96  & $-$3.71  &  10.5  & 0.15 & 0.320  & 0.290 \\
SPT-S J030956-6058.6 & 150 & 59525 & 59537 &  90   &  7.50  &  4.24  & 0.599 & 0.0591 & 0.0134 \\
SPT-S J013305-5200.1 & 150 & 59889 & 59904 & 180   & 12.6   & $-$8.13  & 0.595 & 0.146  & 0.0453 \\
\enddata
\tablecomments{Table columns are as follows: (1) SPT identifier, (2) frequency band, (3,4) start and end times of the EVPA swing, (5) EVPA swing amplitude, (6) rate, (7) observed time lag, (8) peak flux of associated flare, (9) probability of coincidental association, (10) probability of coincidental association with a flare of height $\geq F_{\text{peak}}$.}
\end{deluxetable*}

\begin{figure}[th]
    \plotone{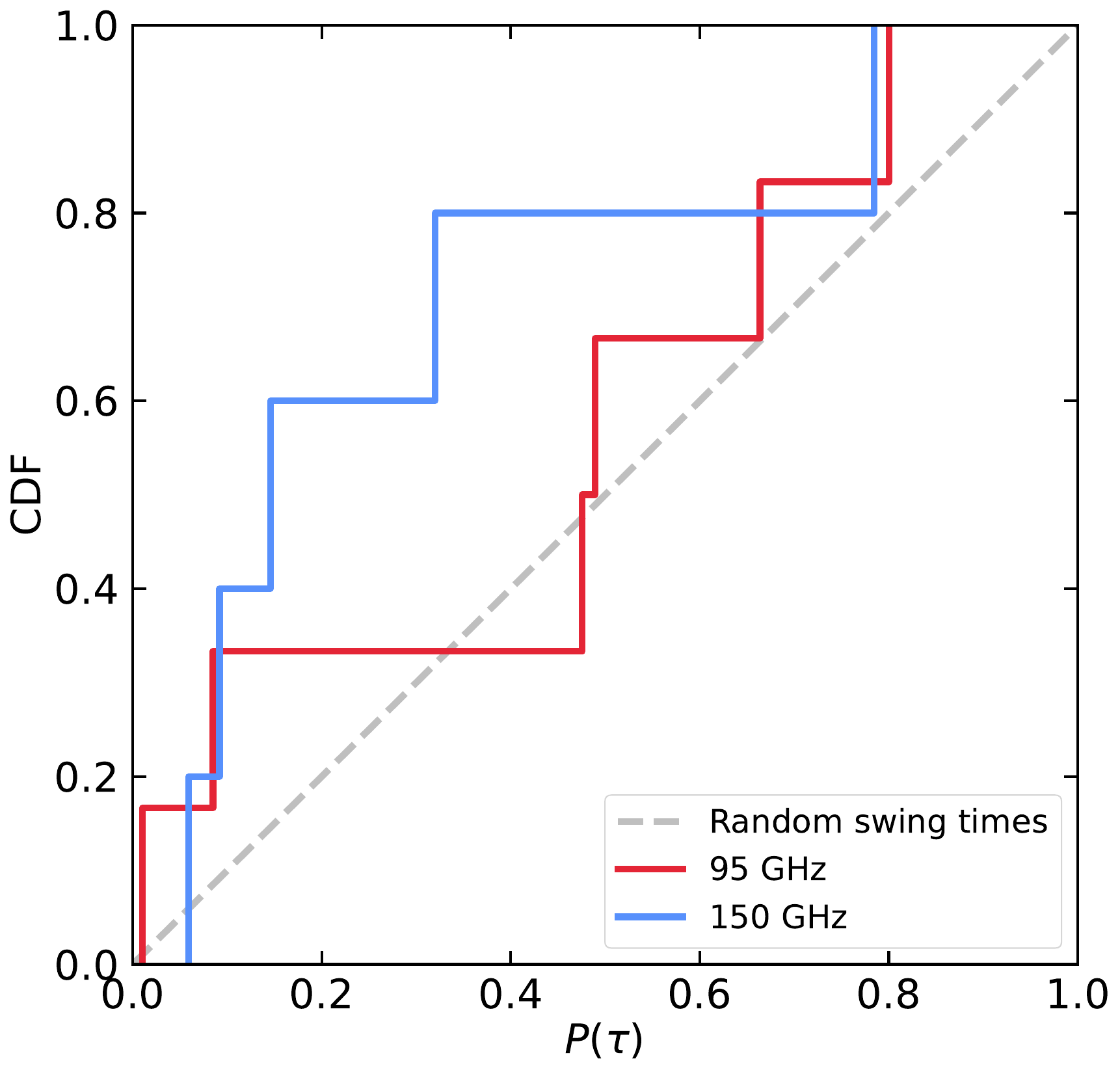}
    \caption{
    CDF of the probability of accidental association between gamma-ray flares and all EVPA swings in 95~GHz (solid, red), 150~GHz (solid, blue), and the simulated random EVPA swings (dashed, gray).
    While some EVPA swings individually have low probability of coincidental association, the time lag distributions of 95 and 150~GHz EVPA swing associated with gamma-ray flares are not significantly different from the randomly selected EVPA swing times (A-D p-values of $>0.25$ and $0.11$ respectively). 
    }
    \label{fig:Ptau_cdf}
\end{figure}

The probability of chance association with any flare, and with a flare of peak height greater than or equal to the observed associated flare, computed as described in Sec.~\ref{SubSec:Probability of coincidental flare associations}, are recorded in columns 9 and 10 of Table~\ref{table:gamma-ray_swings}. The P$(\tau)$ values range from $0.00103$ to $0.489$, while the P($\tau$,$F_{\text{peak}}$) values range from $0.000614$ to $0.447$. We perform an A-D test on the P$(\tau)$ values compared to those from the simulations (distributions shown in Fig.~\ref{fig:Ptau_cdf}) and find
that the distributions of all probabilities of both 95 and 150~GHz EVPA swings associated with gamma-ray flares are not significantly different (A-D p-values of $>0.25$ and $0.11$ respectively) than those with randomized EVPA swing times. Thus, while we find some individual swings with very low probabilities of coincidental association, the overall distribution of swings does not appear to be significantly correlated with gamma-ray flares.
We also note that, in each band (95 and 150~GHz), three of the four low-time-lag gamma-ray-associated EVPA swings (see Table~\ref{table:gamma-ray_swings}) have positive time delays (EVPA swing lags gamma-ray flare).

\section{Discussion and conclusion}
\label{Sec:Conclusion}

Following previous studies that have shown large EVPA changes in blazars in the mm-wave on long timescales \citep{agudo22}, and a parsec-scale $90^\circ$ EVPA change in PKS 0446+11 seen by \citet{kovalev26}, we present the first systematic identification of EVPA swings at mm wavelengths, detecting 14 EVPA swings over a period of five years in eight blazars within the SPT-3G Main field.
We take precautions to prevent spurious identification of EVPA swings, including a PD SNR cut on individual EVPA observations, calculation of the probability of correct measurement, and the requirement that EVPA swings have greater than two data points ($N > 2$) and the number of significant changes is greater than one ($N_{\text{sig}} > 1$).
We do acknowledge that, at least partially owing to the relatively short history of this field of investigation, many of variables used in this study (EVPA swings, gamma-ray flares, time delays) are defined somewhat arbitrarily.

It is possible that some EVPA swings are missed by our analysis due to a decrease in polarized flux, causing individual observations to fall below the PD SNR threshold of three and be removed from the EVPA light curve.
In these cases, missing observations can cause the EVPA swing to have smaller amplitudes, due to missing start or end points, or to be missed entirely.
This is a potential cause of the discrepancy between the 95 and 150~GHz results, particularly the difference in significance of gamma-ray flare associations and number of EVPA swings between the two bands. 
The average PD at 95~GHz is systematically lower than at 150~GHz for most sources in our sample, resulting in higher per-observation EVPA uncertainties and a larger fraction of data removed by the SNR cut when PD decreases significantly.
This is a potential cause of instances where a 150~GHz EVPA swing is detected while no EVPA swing is detected at 95~GHz.

The 14 EVPA swings we identify are detected only within sources in our gamma-ray subsample, consistent with optical studies \citep[e.g.,][]{blinov15, blinov16}, and suggesting that the physical process driving EVPA swings is connected to high-energy emission.
The mean amplitude of our observed EVPA swings ($124^\circ$) is somewhat smaller than that found by \citet{blinov15} ($\approx 180^\circ$), and the mean rotation rate ($5.60^\circ$/day) is similarly lower than the optical mean of $\approx 12.2^\circ$/day. 
This is consistent with the expectation that mm-wave EVPA variability is slower than at optical wavelengths \citep[e.g.,][]{agudo22}, reflecting the more extended and less rapidly evolving magnetic field structure of the mm-wave emitting region further downstream in the jet.

We find a wide range of EVPA variability across the Main Sample on one-week timescales.
SPT-S~J205616-4714.8 is a clear outlier, showing significantly higher variability than all other sources, while also having the lowest average PD in the sample ($2.00\%$ at 95~GHz and $2.43\%$ at 150~GHz respectively). 
In contrast, the two of the three sources with the lowest variability, SPT-S~J025329-5441.8 and SPT-S~J231545-5018.6, have the highest PDs in the Main Sample ($5.72\%$ and $6.50\%$ in 95~GHz and $6.08\%$ and $6.57\%$ in 150~GHz, respectively).
This anti-correlation between PD and rotation rate variability has been previously observed in optical studies \citep[e.g.,][]{blinov16, otero-santos23}, and is also reflected in the visual inspection of Fig.~\ref{fig:gamma-ray_LCs}, where many EVPA swings are accompanied by a significant decrease in PD.
These results are consistent with models in which sources with more ordered magnetic fields exhibit less erratic EVPA variability, and in which EVPA swings involve a temporary disruption or reorientation of an ordered field structure.

Comparing the observed EVPA swing properties to random walk simulations, we find that the random walk simulations do not reproduce the distribution of EVPA swing amplitudes (A-D p-value $=0.035$), durations (A-D p-value $=0.0066$), and rotation rates (A-D p-value $<0.001$).
More decisively, the simulations produce significantly fewer EVPA swings per blazar than observed (A-D p-value $< 0.001$).
We note that the pure random walk tested here is the simplest version of the stochastic model.


Despite the fact all 14 identified EVPA swings are from sources in our gamma-ray sub-sample (with a $<1\%$ probability of happening by chance), and the fact that we find individual swing-flare associations with a very low probability of being coincidental, we find no significant difference between the full distribution of observed time delays between 95 and 150~GHz EVPA swings and gamma-ray flares and those found in simulations with no true assocation (A-D p-values of $>0.25$ and $0.11$ for 95 and 150~GHz respectively).
There are several ways to interpret these somewhat contradictory results, one of which is that our detected EVPA swings come in two types, only one of which is associated with gamma-ray flares, and the fact that the other type also only occurs in the gamma-ray subsample of this analysis is a chance coincidence.
Another is that all swings are truly associated with gamma-ray flaring activity, but that the associated flare can be missed in \textit{Fermi} data, either because it is not strong enough to be detected by our flare finder, or, more speculatively, from a combination of misalignment between the flare emission and the line of sight and different levels of relativistic beaming between the gamma-ray and mm-wave emission, as in proposed orphan GRB afterglows \citep[e.g.,][]{piran04}.

Even in the scenario in which at least some of the EVPA swings in our dataset are truly physically associated with gamma-ray flares, using these associations to identify the emission mechanisms for these EVPA swings remains challenging with our dataset.
\citet{bolis26} show that a purely geometric and deterministic jet model can produce complex EVPA variability on intra-day timescales that our $\sim$2-day cadence cannot resolve.
With our $\sim$2-day cadence, limited number of observed EVPA swings, and lack of multi-wavelength polarization data (e.g. optical or x-ray), we cannot distinguish models that predict intra-day or frequency dependent EVPA variability \citep[e.g.,][]{zhang15}.


A catalog of blazar light curves observed in SPTpol \citep{hood26} is already publicly available, and the SPT-3G light curve catalog, which will include the sources studied here along with the other 165 bright sources within the 1500~deg$^2$ SPT-3G Main field, will be made available in the near future.
The blazars in our Main Sample will continue to be monitored as the SPT-3G survey continues to observe this patch of sky, extending the five-year baseline presented here and potentially increasing the number of detected EVPA swing events.
This continued monitoring, combined with similar wide-field mm-wave polarization programs such as the Simons Observatory \citep{ade19, abitbol25}, which will survey a significantly larger fraction of the sky with similar (but wider) frequency coverage and comparable per-observation depth, will provide yet larger and more diverse blazar samples. 
These samples have the potential to give the results presented here more statistical significance and to more fully characterize the connection between mm-wave EVPA swings and gamma-ray flaring activity in blazar jets.

\begin{acknowledgments}
The South Pole Telescope program is supported by the National Science Foundation (NSF) through awards OPP-1852617 and OPP-2332483. Partial support is also provided by the Kavli Institute of Cosmological Physics at the University of Chicago.
ACT-related work was supported by the U.S. National Science Foundation through awards AST-0408698, AST-0965625, and AST-1440226 for the ACT project, as well as awards PHY-0355328, PHY-0855887, and PHY-1214379. Funding was also provided by Princeton University, the University of Pennsylvania, and a Canada Foundation for Innovation (CFI) award to UBC.
ACT operated in the Parque Astron\'omico Atacama in northern Chile under the auspices of the Agencia Nacional de Investigaci\'on y Desarrollo (ANID; formerly Comisi\'on Nacional de Investigaci\'on Cient\'ifica y Tecnol\'ogica de Chile, or CONICYT). We thank the Republic of Chile for hosting ACT in the northern Atacama, and the local indigenous Licanantay communities whom we follow in observing and learning from the night sky.
Detector research at NIST was supported by the NIST Innovations in Measurement Science program. Computing for ACT was performed using the Princeton Research Computing resources at Princeton University, the National Energy Research Scientific Computing Center (NERSC), and the Niagara supercomputer at the SciNet HPC Consortium. SciNet is funded by the CFI under the auspices of Compute Canada, the Government of Ontario, the Ontario Research Fund--Research Excellence, and the University of Toronto.
Colleagues at AstroNorte and RadioSky provided logistical support for ACT and kept operations in Chile running smoothly. We also thank the Mishrahi Fund and the Wilkinson Fund for their generous support of the project.
Argonne National Laboratory’s work was supported by the U.S. Department of Energy, Office of High Energy Physics, under contract DE-AC02-06CH11357. 
The UC Davis group acknowledges support from Michael and Ester Vaida. 
Work at the Fermi National Accelerator Laboratory (Fermilab), a U.S. Department of Energy, Office of Science, Office of High Energy Physics HEP User Facility, is managed by Fermi Forward Discovery Group, LLC, acting under Contract No. 89243024CSC000002.
The Melbourne authors acknowledge support from the Australian Research Council’s Discovery Project scheme (No. DP210102386). 
The Paris group has received funding from the European Research Council (ERC) under the European Union’s Horizon 2020 research and innovation program (grant agreement No 101001897), and funding from the Centre National d’Etudes Spatiales. 
The SLAC group is supported in part by the Department of Energy at SLAC National Accelerator Laboratory, under contract DE-AC02-76SF00515.
This research was done using services provided by the OSG Consortium \citep{pordes07, sfiligoi09, osg06, osg15}, which is supported by the National Science Foundation awards \# 2030508 and \# 2323298.
Some results in this paper have been derived using \citet{2024ascl.soft02006K}. 
A.S. also acknowledges partial support from the GEM fellowship program.
ADH acknowledges support from the Sutton Family Chair in Science, Christianity and Cultures, from the Faculty of Arts and Science, University of Toronto, from the Natural Sciences and Engineering Research Council of Canada (NSERC) [RGPIN-2023-05014, DGECR-2023-00180].
\end{acknowledgments}

%
\facilities{SPT(SPT-3G), ACT, FGST}



\bibliography{spt_before_1995, spt_1995_to_2000, spt_2000_to_2005,spt_2005_to_2010, spt_2010_to_2015, spt_2015_to_2020, spt_2020_to_2025, spt_2025_and_after}{}
\bibliographystyle{aasjournalv7}

\appendix

\section{Five-year light curves of main sample blazars}

\begin{figure*}[htbp]
\plotone{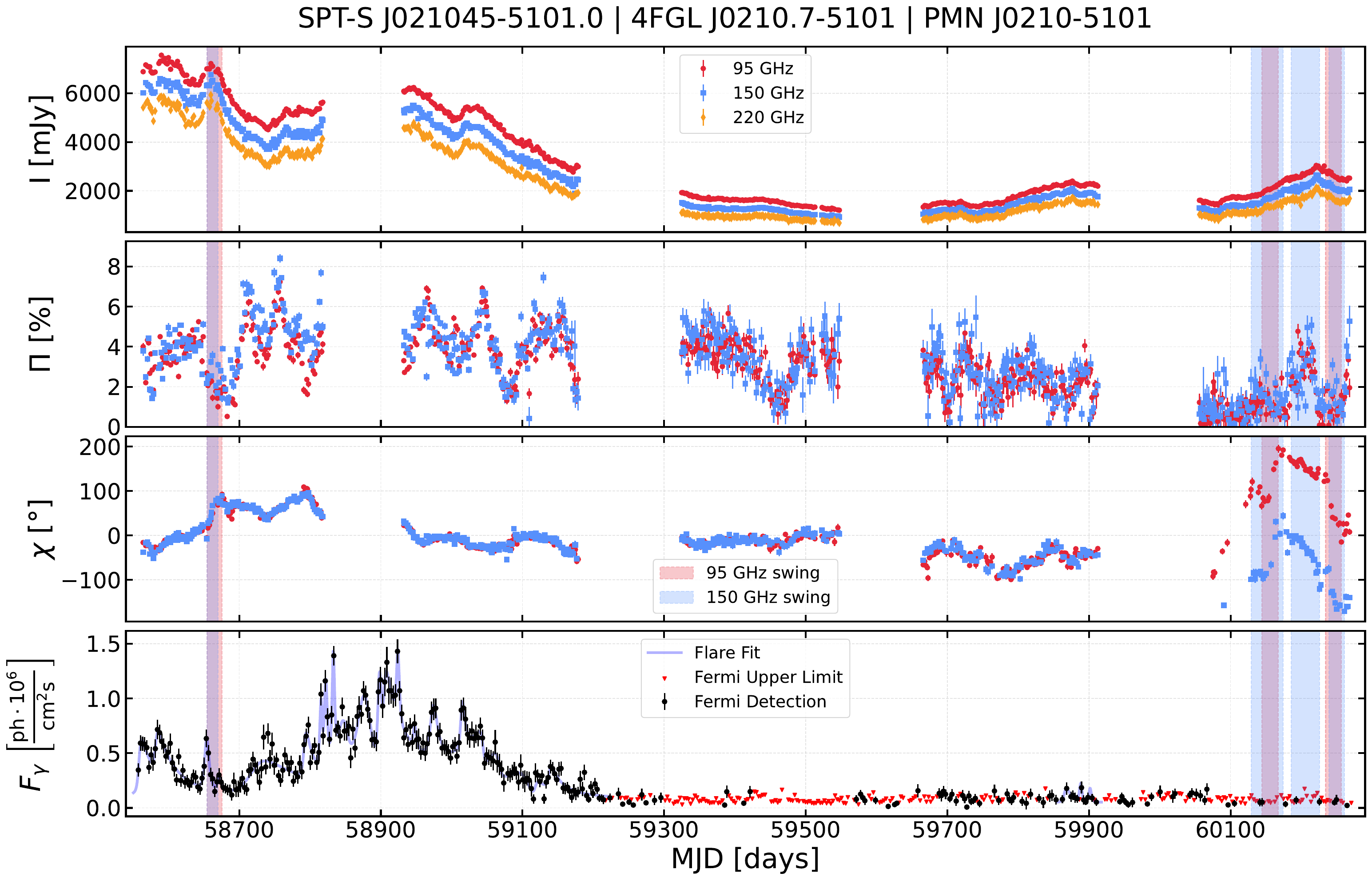}
\caption{Five-year light curves for the gamma-ray subsample of blazars. In each panel, from top to bottom, the rows are as follows: (1) Flux density in mJy for the 95 (red circle), 150 (blue square), and 220 (orange diamond) GHz frequency bands, (2) PD ($\Pi$), (3) adjusted EVPA ($\chi$), (4) \textit{Fermi} gamma-ray $3\sigma$ detection flux (black), upper limits (red triangle), and flare fit (blue), identified swings in the 95 (red highlight) and 150~GHz bands (blue highlight).\label{fig:gamma-ray_LCs}}
\end{figure*}

\begin{figure*}[htbp] \ContinuedFloat
\plotone{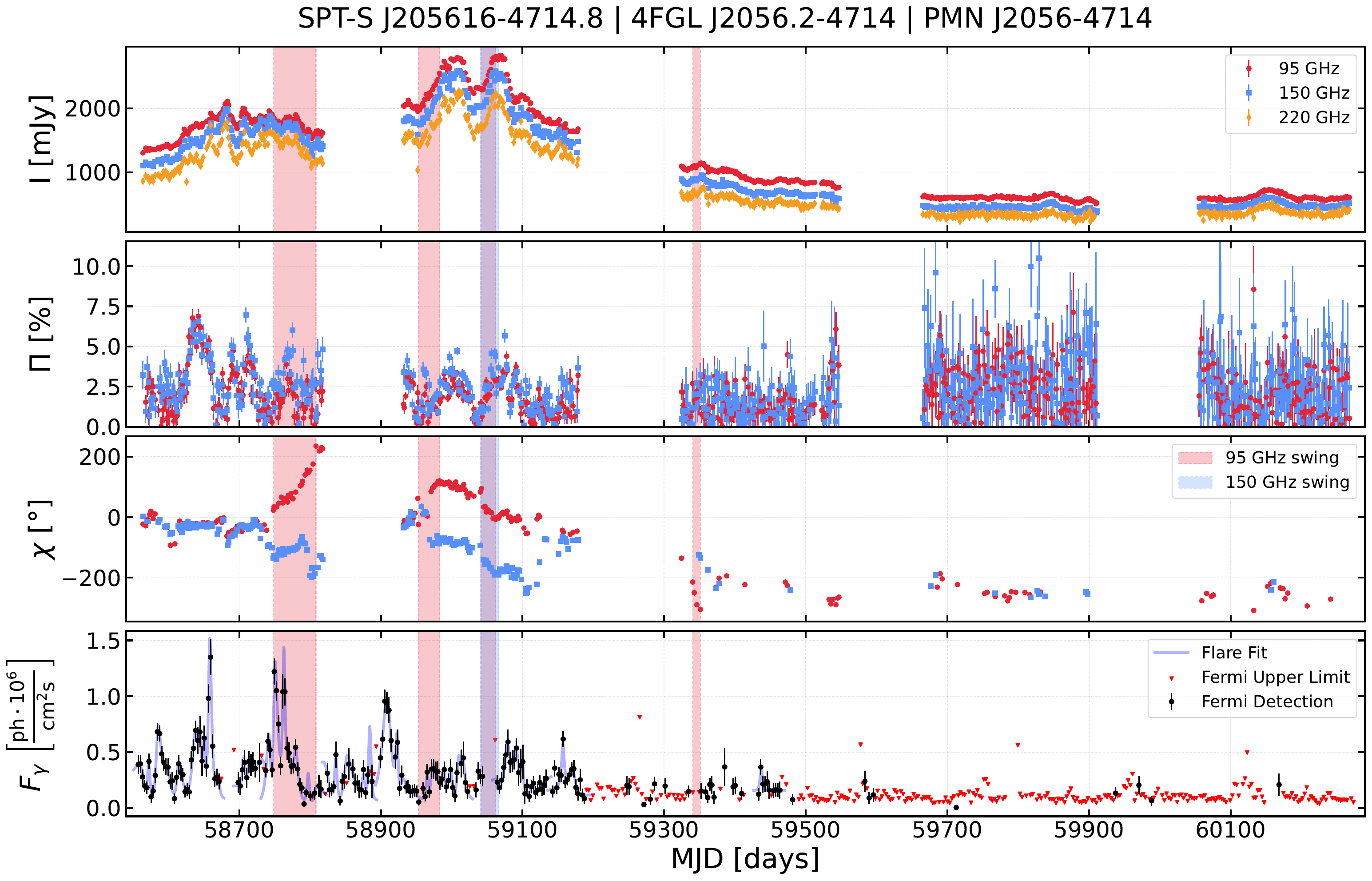}
\caption{continued.}
\label{cont2}
\end{figure*}

\begin{figure*}[htbp] \ContinuedFloat
\plotone{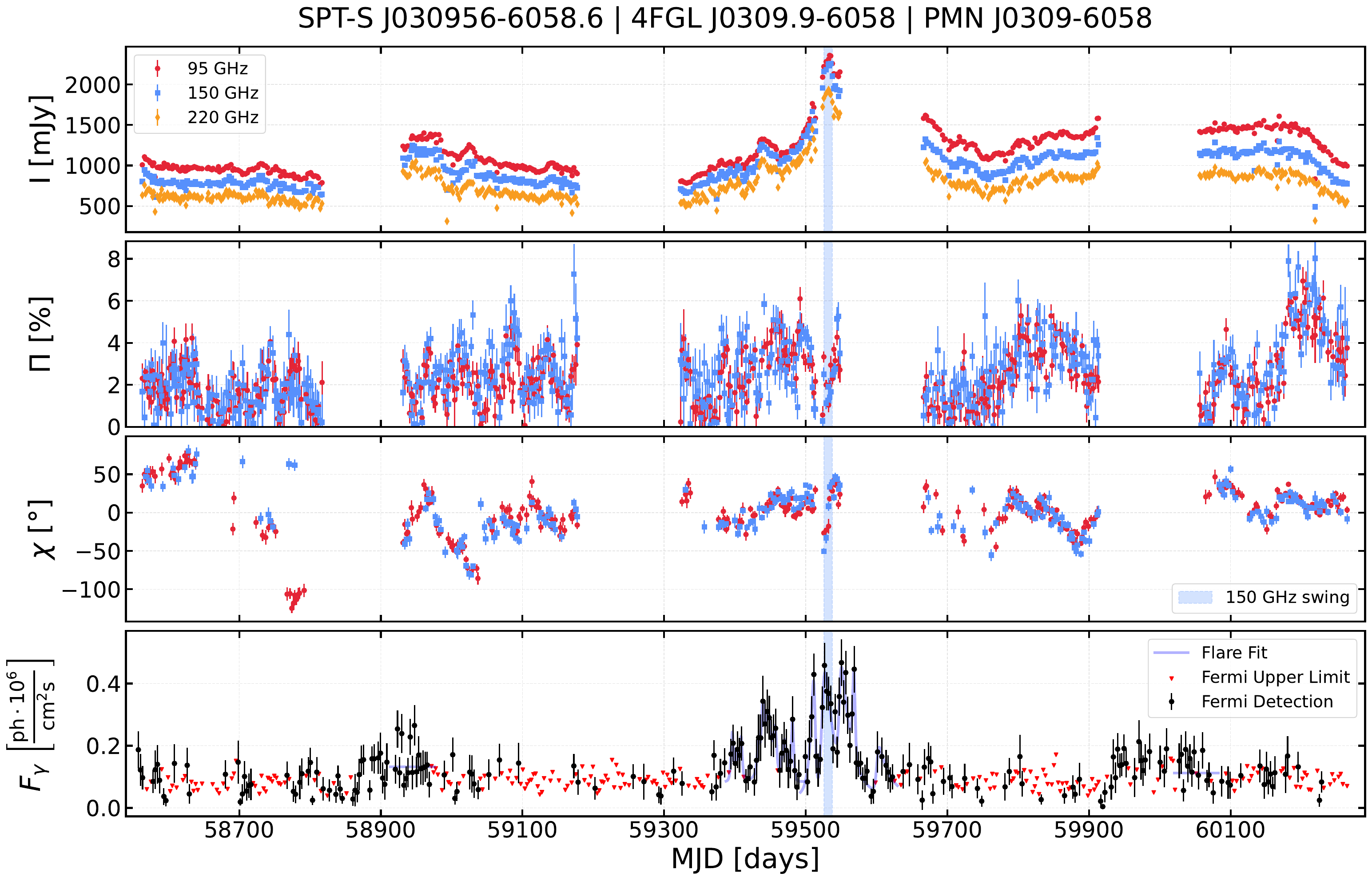}
\caption{continued.}
\label{cont3}
\end{figure*}

\begin{figure*}[htbp] \ContinuedFloat
\plotone{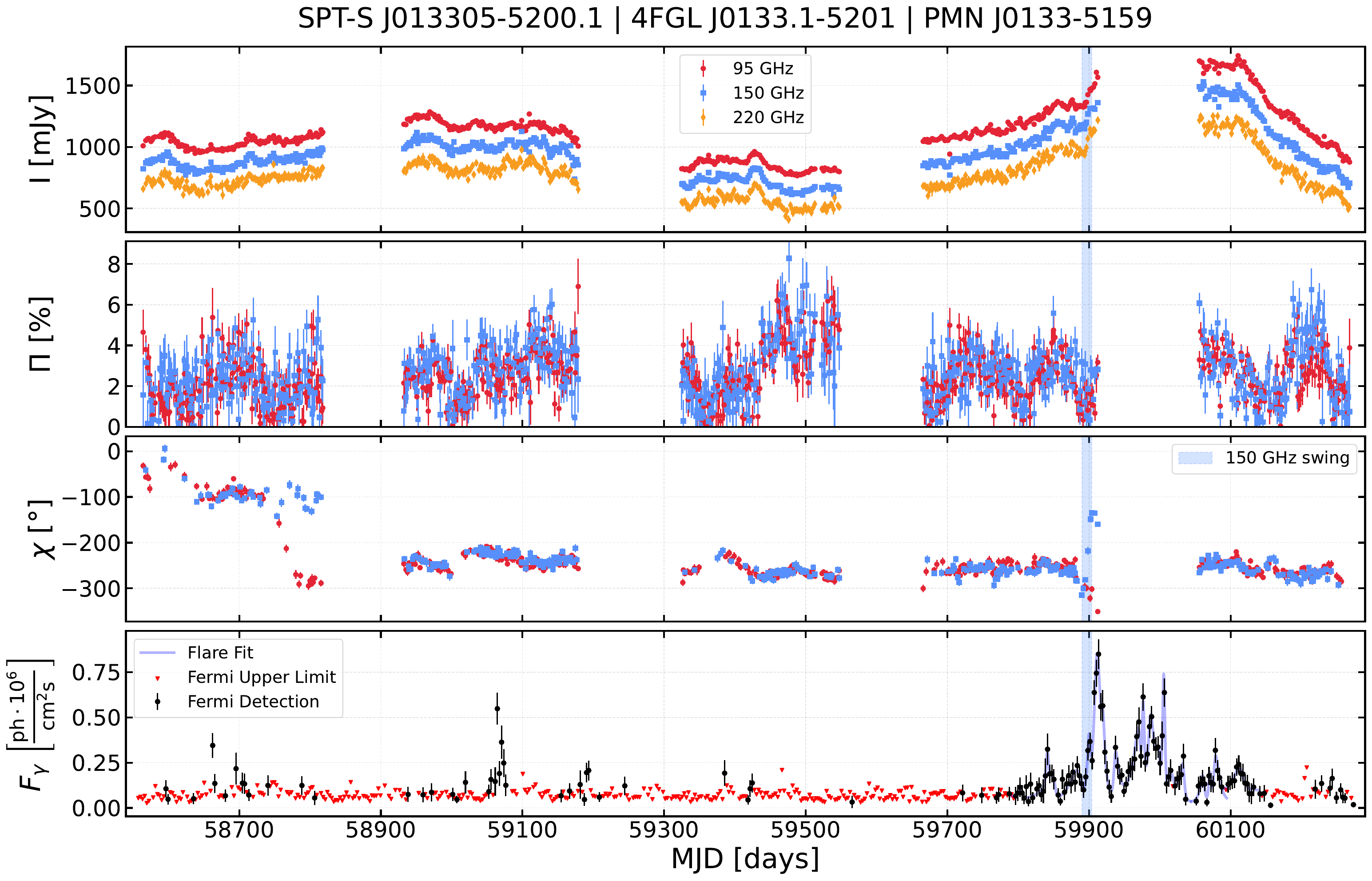}
\caption{continued.}
\label{cont4}
\end{figure*}

\begin{figure*}[htbp]
\plotone{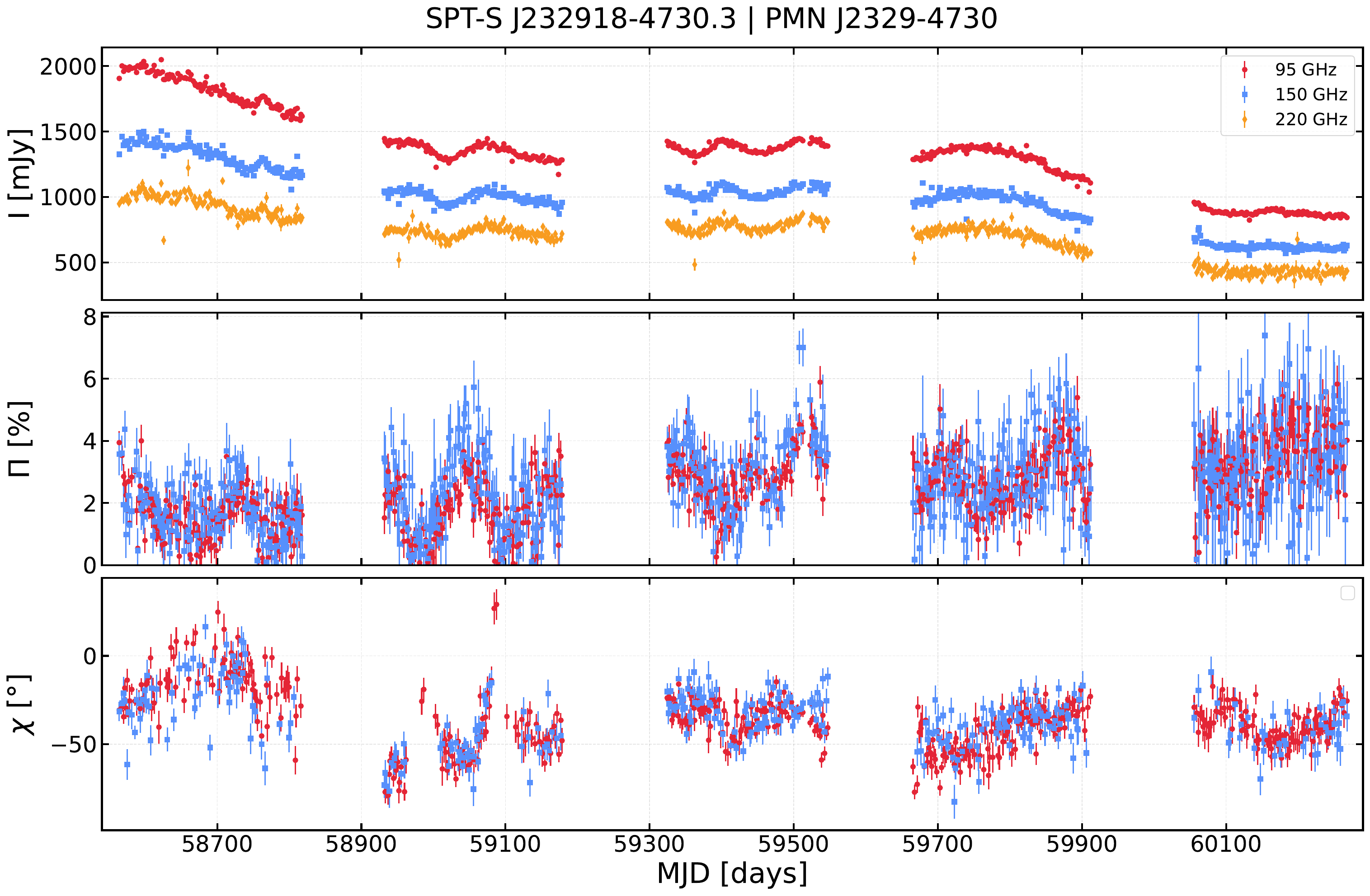}
\caption{Five-year light curves for the non-gamma-ray subsample of blazars. In each plot, from top to bottom, the rows are as follows: (1) Flux density in mJy for the 95 (red circle), 150 (blue square), and 220 (orange diamond) GHz frequency bands, (2) PD ($\Pi$), (3) adjusted EVPA ($\chi$), identified swings in the 95 (red highlight) and 150~GHz bands (blue highlight).\label{fig:non_gamma-ray_LCs}}
\end{figure*}

\begin{figure*}[htbp] \ContinuedFloat
\plotone{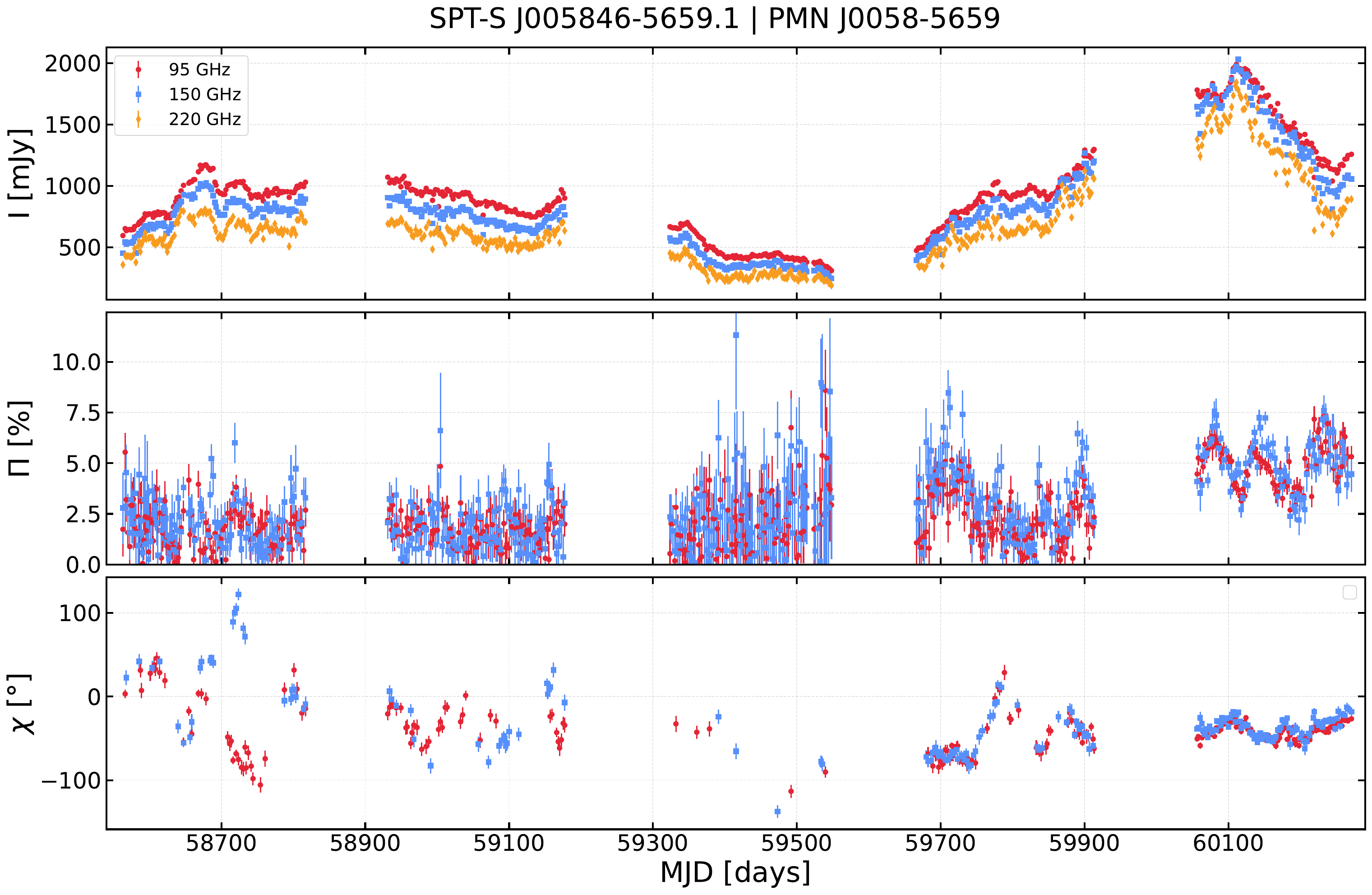}
\caption{continued.}
\end{figure*}

\begin{figure*}[htbp] \ContinuedFloat
\plotone{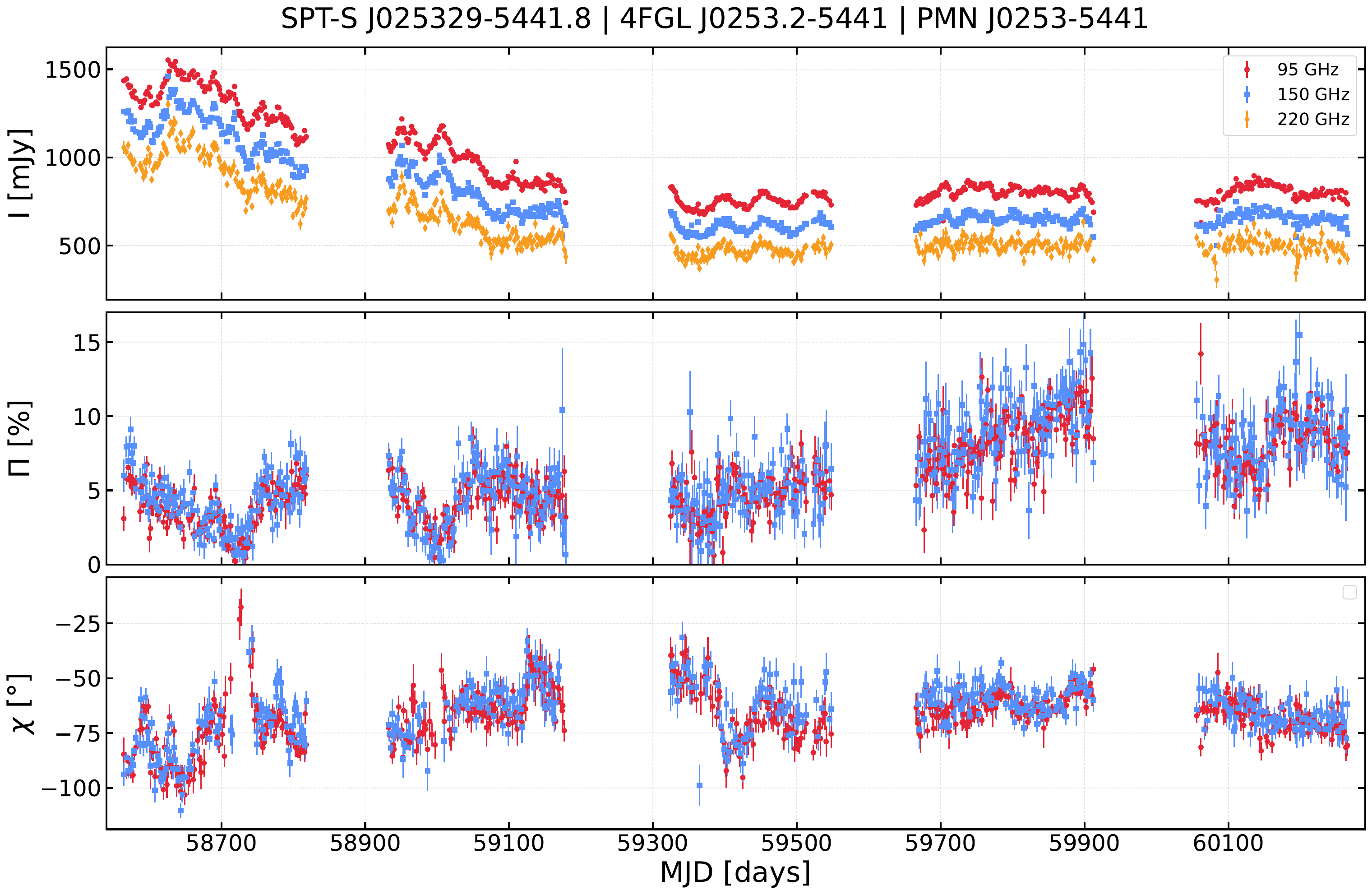}
\caption{continued.}
\end{figure*}

\begin{figure*}[htbp] \ContinuedFloat
\plotone{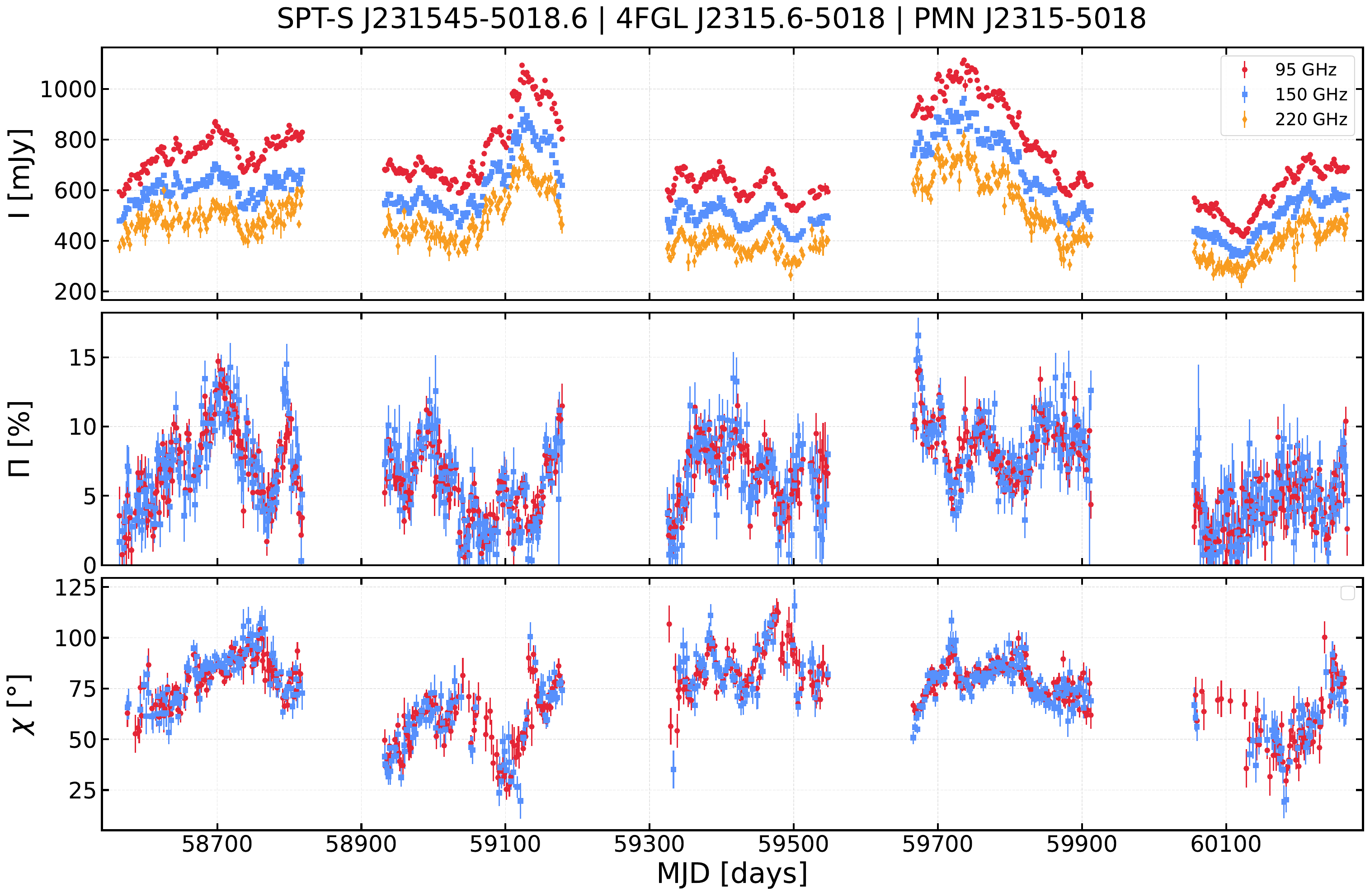}
\caption{continued.}
\end{figure*}



\end{document}